\documentclass[ reprint, twocolumn, superscriptaddress, amsmath,amssymb, aps,prb]{revtex4-2}

\usepackage{graphicx}
\usepackage{dcolumn}
\usepackage{bm}
\usepackage{lmodern}
\usepackage[usenames,dvipsnames]{color}
\usepackage{tablefootnote}
\usepackage{xcolor}
\usepackage[normalem]{ulem}
\usepackage[colorlinks = true,
            linkcolor = Black,
            urlcolor  = Blue,
            citecolor = Black,
            anchorcolor = Blue]{hyperref}

\newcommand{\q}{\mathbf{q}}
\newcommand{\kk}{\mathbf{k}}

\begin{document}


\title{Nonadiabatic reactive scattering of hydrogen on different surface facets of copper} 

\author{Wojciech G. Stark}
\affiliation{Department of Chemistry, University of Warwick, Coventry, CV4 7AL, United Kingdom}
\author{Connor L. Box}%
\affiliation{Department of Chemistry, University of Warwick, Coventry, CV4 7AL, United Kingdom}
\author{Matthias Sachs}%
\affiliation{School of Mathematics, University of Birmingham, Birmingham, B15 2TS, United Kingdom}
\author{Nils Hertl}%
\affiliation{Department of Chemistry, University of Warwick, Coventry, CV4 7AL, United Kingdom}
\affiliation{Department of Physics, University of Warwick, Coventry, CV4 7AL, United Kingdom}
\author{Reinhard J. Maurer}%
 \email{r.maurer@warwick.ac.uk}
\affiliation{Department of Chemistry, University of Warwick, Coventry, CV4 7AL, United Kingdom}
\affiliation{Department of Physics, University of Warwick, Coventry, CV4 7AL, United Kingdom}

\date{\today}

\begin{abstract}
Dissociative chemisorption is a key process in hydrogen-metal surface chemistry, where nonadiabatic effects due to low-lying electron-hole-pair excitations may affect reaction outcomes. Molecular dynamics with electronic friction simulations can capture weak nonadiabatic effects at metal surfaces, but require as input energy landscapes and electronic friction tensors. Here, we present full-dimensional machine learning surrogate models of the electronic friction tensor to study reactive hydrogen chemistry at the low-index surface facets Cu(100), Cu(110), Cu(111), and Cu(211). We combine these surrogate models with machine learning interatomic potentials to simulate quantum-state-resolved H$_2$ reactive scattering on pristine copper surfaces. The predicted sticking coefficient and survival probabilities are in excellent agreement with experiment. Comparison between adiabatic and nonadiabatic simulations reveals that the influence of electron-hole pair excitations on the scattering dynamics is weak and that the probability for dissociative adsorption is dominated by the shape of the underlying potential energy surface and the initial vibrational quantum state. Nonadiabatic effects only lead to subtle changes in rovibrationally inelastic state-to-state scattering probabilities. The differences between jellium-based isotropic local density friction and full ab-initio response-theory-based orbital-dependent friction are even more subtle. The presented machine learning models represent the most accurate publicly available, full-dimensional models for H$_2$ on copper to date.
 \end{abstract}
\maketitle


\section{Introduction} \label{sec:intro}

Dissociative adsorption of hydrogen on metal surfaces plays a vital role in many crucial processes, such as heterogeneous catalysis~\cite{andersson_structure_2008}, production of molecular hydrogen~\cite{hollenbach_surface_1971}, olefin hydrogenation~\cite{wilde_influence_2008}, hydrogen storage~\cite{lee_hydrogen_2000}, and many others~\cite{kirchheim_hydrogen_1988, higashi_ideal_1990, haberer_tunable_2010, liu_roles_2016}. At the atomic level, these macroscopic processes involve a complex reaction network, rendering the detailed characterization of individual elementary steps highly  challenging~\cite{ertl_reactions_2008}. 

To better isolate specific elementary steps in gas-surface dynamics, experiments and simulations are commonly performed at idealized single-crystal surfaces. The incoming gaseous particles are prepared in well-defined vibrational and rotational quantum states. In combination with quantum-state selective measurement techniques, such as resonance enhanced multiphonon ionization (REMPI), and angular-resolved time-of-flight measurements, it is possible to realize detailed state-to-state scattering experiments \cite{auerbach_chemical_2021}. These experiments are useful in scrutinizing the scattering dynamics of the impinging molecules and the concomitant energy transfer from the molecule to the surface, as well as the intramolecular energy redistribution. The dynamics of hydrogen molecules on metal surfaces are particularly intriguing: Many metal catalysts readily catalyze H\textsubscript{2} dissociation, enabling the catalytic processes outlined above.~\cite{farias_diffraction_2011}  The light mass of hydrogen means that nuclear quantum effects likely also play a role in gas-surface dynamics.~\cite{zhang_ring_2025} Additionally, for metallic substrates, the scattering process can be affected by nonadiabatic effects and energy dissipation processes that arise from the excitation of electron-hole pairs (ehps) in the metal. These effects have the potential to modify reaction outcomes.~\cite{maurer_mode_2017,maurer_hot_2019, spiering_testing_2018}  Hence, quantum-state resolved H\textsubscript{2} scattering from pristine metal surfaces constitutes an ideal stress test for existing concepts in quantum mechanics and theoretical gas-surface dynamics and the prediction capabilities of first principles simulation methods.~\cite{kroes_computational_2021}

Over the last decades, H\textsubscript{2} on copper surfaces has become a commonly studied model system when it comes to quantum-state resolved dissociative adsorption \cite{michelsen_state-specific_1992, rettner_dynamics_1992, rettner_determination_1993, watts_state--state_2001, shackman_state--state_2005, hodgson_scattering_1991,  murphy_adsorption_1998, cao_hydrogen_2018} and the concomitant kinetics \cite{anger_adsorption_1989, rasmussen_dissociative_1993, campbell_the_1991, nitz_thermal_2024} as well as associative desorption dynamics of diatomic molecules from metal surfaces \cite{michelsen_critical_1991, michelsen_on_1992, michelsen_effect_1993, rettner_quantumstatespecific_1995, kaufmann_associative_2018, murphy_inverted_1999}. A key finding that emerged from the vast literature is that vibrational excitation of H\textsubscript{2} significantly promotes its dissociation over copper. The extent to which vibrational excitation promotes surface dissociation reactions is typically quantified by the vibrational efficacy. \cite{diaz_note_2009} This quantification attempt took inspiration from the well-known Polanyi rules \cite{polanyi_concepts_1972}, which connect the ability to overcome the reaction barrier through vibrational excitation with the shape of the reaction pathway for reactions of the type
A+BC$\rightarrow$AB+C: Reactions with barriers that appear early on the pathway can be promoted by increasing the translational energy. Reactions with late barriers, on the other hand, are promoted by vibrational excitations. 



While most computational studies of H\textsubscript{2} interacting with copper surfaces have been performed within the Born–Oppenheimer approximation (BOA), this framework assumes that the nuclear dynamics evolve solely on the ground-state potential energy surface (PES). The BOA generally holds for systems with well-separated electronic states; however, at metal surfaces, particularly under vibrational excitation, this assumption does not generally hold. This is well documented for systems such as NO and CO scattering from Au(111) and Ag(111), where electronic excitations can substantially influence reaction dynamics \cite{huang_vibrational_2000, kruger_vibrational_2016, shenvi_nonadiabatic_2009, meng_first-principles_2024}.

For H\textsubscript{2} on copper, the possibility of energy dissipation into ehp excitations has motivated studies beyond the BOA, particularly using molecular dynamics with electronic friction (MDEF).\cite{headgordon_molecular_1995} Within this framework, most previous work on gas-surface scattering has relied on the local density friction approximation (LDFA) \cite{li_nonadiabatic_1992, juaristi_role_2008}, which models friction as an isotropic function of the local electron density. While LDFA-based simulations generally report only a weak impact of nonadiabatic effects on the sticking probability of H\textsubscript{2} on copper \cite{juaristi_role_2008, muzas_vibrational_2012}, the method’s applicability to molecular scattering has been questioned—both for its reliance on isotropic damping and for its grounding in jellium-based models rather than ab initio electronic structure calculations \cite{luntz_comment_2009, box_determining_2021}.

Electronic friction tensors (EFTs) have also been calculated directly from density functional theory (DFT) via first-order time-dependent perturbation theory (TDPT).~\cite{hellsing_electronic_1984, headgordon_molecular_1995, luntz_comment_2009} This approach is also called orbital-dependent friction (ODF). The resulting friction tensors are fully anisotropic and capture coordinate and mode-dependent differences in nonadiabatic relaxation during atom and molecule dynamics at surfaces,~\cite{askerka_role_2016, maurer_ab_2016, box_ab_2023, spiering_orbital-dependent_2019} including hydrogen atom scattering~\cite{box_room_2024}.

Spiering and Meyer \cite{spiering_testing_2018} identified vibrational de-excitation from $\text{H}_2(\nu=2, J=1)\rightarrow\text{H}_2(\nu=1, J=1)$ of H\textsubscript{2} on Cu(111) as an observable that may be particularly sensitive to nonadiabatic effects. Their comparison between LDFA and ODF revealed a distinct kinetic energy dependence of vibrational de-excitation probability that only arises when full tensorial and anisotropic electronic friction effects are considered. However, their model was restricted to six dimensions, one surface facet, and neglected thermal surface motion, limiting its generality.

Recent advances in machine learning interatomic potentials (MLIPs) enable the efficient simulation of gas-surface reaction probabilities, considering all degrees of freedom, including surface phonons.~\cite{behler_generalized_2007, behler_representing_2014, deringer_machine_2019, kocer_neural_2022} Such models, among several other systems, have recently been proposed for a variety of systems, including reactive H\textsubscript{2} scattering on various facets of copper \cite{zhu_unified_2020, stark_machine_2023, stark_benchmarking_2024}. To efficiently perform MDEF simulations with MLIPs, high-dimensional configuration-dependent representations of the EFT are required. Such representations have been recently proposed~\cite{zhang_symmetry-adapted_2020, sachs_machine_2025}  and applied to systems such as NO on Au(111).~\cite{box_determining_2021} In particular, the general model proposed by Sachs \textit{et al.}~\cite{sachs_machine_2025} is appealing as it is based on a rotationally equivariant Atomic Cluster Expansion \cite{drautz_atomic_2019, drautz_atomic_2020} construction that is intrinsically transferable and size extensible, enabling the representation and evaluation of friction across configurational space. With the emergence of the above-mentioned ML surrogate models, comparative studies of nonadiabatic surface chemistry across different surface facets become possible.

In this work, we present high-dimensional machine learning surrogate models of EFTs for H\textsubscript{2} on four surface facets of copper based on LDFA and ODF friction. Upon careful validation of the EFT models, we combine them with a recently developed transferable MLIP for reactive surface chemistry of H\textsubscript{2} on copper~\cite{stark_machine_2023, stark_benchmarking_2024} to perform nonadiabatic MDEF simulations of H\textsubscript{2} state-to-state scattering across four surface facets of copper: Cu(111), Cu(100), Cu(110), Cu(211). The models accurately account for phononic and electronic energy dissipation effects during scattering and achieve excellent agreement with previously reported experimental sticking and survival probabilities of the initial quantum states across different facets. The presented and provided models therefore represent the, to date, most accurate PES and EFT models for reactive hydrogen chemistry on copper. Comprehensive scattering simulations reveal that the impact of electronic nonadiabaticity on reaction probabilities is weak across all surface facets, which is in line with previous findings \cite{wijzenbroek_dynamics_2016, muzas_vibrational_2012, juaristi_role_2008}. Notably, only a weak kinetic energy dependence of the vibrational de-excitation from $\text{H}_2(\nu=2, J=1)\rightarrow\text{H}_2(\nu=1, J=1)$ is found, with nonadiabatic effects only changing absolute populations by a fraction of a percent. Thus, a measurable signature of nonadiabaticity is unlikely to be found on Cu(111) or any one of the other studied low-index surface facets. 

\section{Methods} \label{sec:methods}

    \subsection{Molecular Dynamics with Electronic Friction} \label{sec:methods_mdef}
        
        The motion of atoms in molecular dynamics with electronic friction (MDEF) follows the following Langevin equation,
        \begin{equation} \label{eq:2_mdef}
             M_i\Ddot{R}_i = - \frac{\partial V (\bm{R},\bm{z})}{\partial R_i} - \sum _j \Lambda_{ij}(\bm{R},\bm{z}) \dot{R}_j + F^{\mathcal{R}}_{i} (t) ,
        \end{equation}
        where the atomic coordinate, $i$, has mass, $M_i$ and position $R_{i}$. 
        $V(\bm{R},\bm{z})$  is the potential energy of the simulated atomistic system. $\Lambda_{ij}(\bm{R},\bm{z})$ is a component of of the $3N\times3N$ EFT $\bm{\Lambda}$, where $N$ is the number of atoms. Both these quantities are functions of the atomic positions $\bm{R} = (R_1,\dots,R_N)$ and their corresponding atomic element types $\bm{z} = (z_1,\dots,z_N)$. $F^{\mathcal{R}}_{i} (t)$ is a white noise force governed by electronic friction.
        
        Maurer~\textit{et~al.}~\cite{maurer_ab_2016} proposed an expression to evaluate $\bm{\Lambda}$ based on first-order TDPT within DFT,
        \begin{equation}\label{eq:odf_sd}
            \begin{split}
                \Lambda_{ij}(\hbar\omega) = & \pi\hbar \sum_{\kk, \sigma, m,n}w_{\kk} \tilde{g}^{i}_{mn} (\kk) [\tilde{g}^{j}_{mn}]^{*} (\kk) \\
                & \times (f_{n \kk}-f_{m \kk}) \frac{\delta(\epsilon_{m \kk}-\epsilon_{n \kk}-\hbar\omega)}{\epsilon_{m \kk}-\epsilon_{n \kk}} ,
            \end{split}
        \end{equation}
        where $ w_{\kk}$ are Brillouin zone sampling weights, $\tilde{g}^{i}_{mn}(\kk)$ are electron-phonon coupling (EPC) matrix elements describing nonadiabatic coupling between Kohn-Sham single particle states $m$ and  $n$ at wave vector $\kk$, and $\sigma$ is the spin label. The states have energies given by $\epsilon_{m \kk}$ and occupations as $f_{n \kk}$. In typical calculations, the EFT is evaluated in the quasi-static limit ($\hbar\omega \to 0$). This formulation of the ODF method accounts for vibrational mode dependence and directional anisotropy of the EFT.
        Also commonly applied is a similar expression that is valid in the low temperature limit (denoted $\bm{\Lambda}^{0}$),\cite{spiering_testing_2018,spiering_orbital-dependent_2019} 
        \begin{equation}\label{eq:odf_dd}
            \begin{split}
                \Lambda_{ij}^{0} = & \pi\hbar \sum_{\kk, \sigma, m,n}w_{\kk} \tilde{g}^{i}_{mn} (\kk) [\tilde{g}^{j}_{mn}]^{*} (\kk) \\
                & \times \delta(\epsilon_{n \kk}-\epsilon_{\mathrm{F}}) \delta(\epsilon_{m \kk}-\epsilon_{\mathrm{F}}) ,
            \end{split}
        \end{equation}
        with $\epsilon_{\mathrm{F}}$ being the Fermi energy. This expression experiences problematic convergence behavior, attributed to the divergence of this expression at $\q=0$.\cite{calandra_electron-phonon_2005, box_ab_2023}
        Box \textit{et al.}~\cite{box_ab_2023} examined the differences between $\Lambda$ and $\Lambda^0$ formulations for H\textsubscript{2} on Cu(111), finding that $\Lambda^0$ yields systematically larger friction values and poorer numerical convergence. As a consequence, $\Lambda^0$ tends to overestimate vibrational lifetimes and phonon linewidths.~\cite{calandra_electron-phonon_2005} 
         We therefore in this work employ \ref{eq:odf_sd} to evaluate the full ODF EFT.

        The \textit{local-density friction approximation (LDFA)} provides an efficient alternative to approximate $\bm{\Lambda}$ by assuming an adsorbate ion interacts with an isotropic, homogeneous electron gas characterized by the local electron density,  $\rho_{\mathrm{emb}}$, of the metal surface: 
        \begin{equation}\label{eq:ldfa}
            \Lambda^{\mathrm{LDFA}}(\rho_{\mathrm{emb}})=\frac{4\pi \rho_{\mathrm{emb}}}{k_{\mathrm{F}}}\sum^{\infty}_{l=1}(l+1)\sin^{2}[\delta_{l}^{F}-\delta^{F}_{l+1}] ,
        \end{equation}
        where $k_{F}$ is the Fermi momentum, and $\delta_{l}^{F}$ is the scattering phase-shift at the Fermi level. Instead of calculating the full EFT, LDFA evaluates an isotropic coefficient that is applied identically along the Cartesian diagonal elements ($\Lambda^{\mathrm{LDFA}}$). This approximation is highly efficient and significantly reduces computational complexity, but neglects the directional dependencies of the dissipation. In practical calculations, the tabulated phase shifts~\citep{puska_atoms_1983,gerrits_electronic_2020} are used, hence, the problem is reduced to evaluating the electron density of the surface metal at the position of the adsorbate atom.

        The EFT can be represented within Cartesian (x, y, z) or internal (d, $\theta$, $\phi$, X, Y, Z) coordinates, as shown in Fig.~\ref{fig:int_cart_coord}. The coordinate transformation is explained in more detail in the Supplemental Material.

        \begin{figure}
            \centering
            \includegraphics[width=2in]{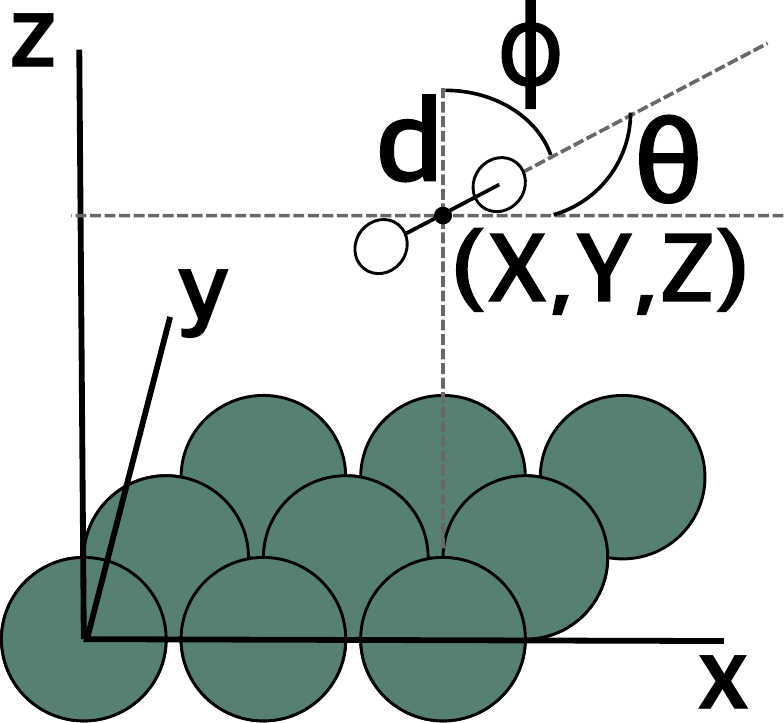}
            \caption{\textbf{Cartesian and internal coordinates for a diatomic molecule.} The Cartesian coordinates (x, y, z) are presented along the internal counterparts (d, $\theta$, $\phi$, X, Y, Z) for H\textsubscript{2} molecule at Cu(111) surface, where d corresponds to an interatomic distance between H atoms, and $\theta$ and $\phi$ to polar and azimuthal angles with respect to the surface normal. X, Y, and Z represent the center-of-mass positions of the adsorbate molecule.}
            \label{fig:int_cart_coord}
        \end{figure}

    \subsection{Machine Learning Representations} \label{sec:methods_ml_eft}

    \paragraph{Atomic Cluster Expansion}
        The ACE~\cite{drautz_atomic_2019} provides a framework to construct highly efficient MLIPs that achieve high body order representation.
        
        In ACE, the atomic basis $A_{z_{i},znlm}$ is defined as a projection of the neighborhood density $\rho_i^{z}$ of all atoms of chemical element $z$ around atom $i$ (where atom $i$ belongs to the chemical element $z_i$) onto the one-particle basis functions $\phi_{nlm}^{z_{i}z_{j}}$
        \begin{equation}
            A_{z_{i},znlm} = \left\langle\rho_{i}^{z}|\phi_{nlm}^{z_{i}z}\right\rangle = \sum_{\substack{j\\
              \forall z_j=z}} \phi_{nlm}^{z_{i}z_{j}}(\bm{r}_{ij})  ,
        \end{equation}
        where  $\bm{r}_{ij} = R_j -R_i$ denotes the displacement of the atom $j$ from the atom $i$, and the one-particle basis functions are summed over atoms $j$, for which $z_{j}=z$. The $z_{i},z_{j}$ are elements $z$ of central $i$ and neighboring $j$ atoms, and the density can be described as
        \begin{equation}
            \rho_{i}^{z}(\bm{r})=\sum_{\substack{j \neq i \\ \forall z_j =z}}\delta(\bm{r}-\bm{r}_{ij}) ,
        \end{equation}
        and the one-particle basis function can be defined as a product of a radial function $R_{nl}^{z_{i}z_{j}}$ and spherical harmonics $Y_{l}^{m}$,
        \begin{equation}
            \phi_{nlm}^{z_{i}z_{j}}(\bm{r})=R_{nl}^{z_{i}z_{j}}(r)Y_{l}^{m}(\bm{\hat{r}}),
        \end{equation}
        where $r = |\bm{r}|$ and $\bm{\hat{r}} = \bm{r}/r$.
        To obtain permutation-invariant basis functions, products of atomic basis functions are constructed 
        \begin{equation}
            A_{z_{i}\bm{\upsilon}}=\prod^{\nu}_{t=1} A_{z_{i}\upsilon_{t}},
        \end{equation}
        where $\upsilon = znlm$, $\bm{\upsilon}=(\upsilon_{1},\dots,\upsilon_{\nu})$ and $\nu$ refers to correlation order, corresponding to a ($\nu$+1)-body basis function. Taking the products of atomic basis functions allows a convenient and efficient construction of basis functions of any body order.
        Isometry invariant basis $B_{z_{i}\eta_{\nu}}$ functions are created from the non-rotationally-invariant  basis functions $A_{z_{i}\bm\upsilon}$ 
        \begin{equation}\label{eq:ACE:scalar}
            B_{z_{i}\eta_{\nu}}=\sum_{\bm{\upsilon}} C_{\eta_{\nu}\bm{\upsilon}}A_{z_{i}\bm{\upsilon}} ,
        \end{equation}
        where $C_{\eta_{\nu}\bm{\upsilon}}$ are generalized Clebsch-Gordan coefficients, and all the allowed invariant contractions of $A_{z_{i}\bm{\upsilon}}$ are enumerated by $\eta_{\nu}$. An interatomic potential model can be defined as a sum of site contributions
        \begin{equation}
            E = \sum_i E_{i}=\sum_i \sum_{\eta_{\nu}} c_{z_{i}\eta_{\nu}}B_{z_{i}\eta_{\nu}} ,
        \end{equation}
        where $c_{z_{i}\eta_{\nu}}$ coefficients are free model parameters optimized in a linear fitting procedure.

    \paragraph{ACE-friction}
    Following the general approach for machine learning of friction tensors introduced in \cite{sachs_machine_2025}, we assume a representation of the friction tensor as 
    \begin{equation}\label{eq:rw-coupling}
          \Lambda_{ij} =
    \sum_{k}{\bf \Sigma}_{ik} \left ({\bf \Sigma}_{jk}\right )^{T}, 
    \end{equation}
    where each ${\bf \Sigma}_{ij}$ is a $3 \times 3$ matrix-valued function 
    of atoms within a pair environment $\mathcal{N}(i,j)$ of the respective atom pair $ij$. That is, for $i\neq j$, the pair environment $\mathcal{N}(i,j)$ contains all atoms within 
    a cutoff region that is rotationally symmetric around the displacement vector ${ R}_{ij}$ and for $i=j$, the environment $\mathcal{N}(i,j)$ reduces to the standard atomic site environment $\mathcal{N}(i)$ defined by a spherical cutoff region centered at ${ R}_i$.  
    Moreover, each ${\bf \Sigma}_{ij}$ is assumed to be matrix-equivariant with respect to the orthogonal group in dimension 3, $O(3)$, in the sense that
    \[
    \forall Q \in O(3): \quad  {\bm \Sigma}_{ij}(Q{\bm R},{\bm z})  = Q{\bm \Sigma}_{ij}({\bm R},{\bm z})Q^{T},
    \]
    and permutation-equivariant with respect to permutations that leave the indices $i,j$ of the respective atom pair unchanged, i.e.,
    \[
    \forall \sigma \in S_N, \sigma i = i, \sigma j = j: ~ {\bm \Sigma}_{ij}(\sigma{\bm R},\sigma{\bm z})  = {\bm \Sigma}_{ij}({\bm R},{\bm z}),
    \]
    where $S_N$ denotes the symmetric group of degree $N$. The action of a  permutation $\sigma \in S_N$ on the position vector, ${\bm R}$, and atomic element vector, ${\bm z}$, is $\sigma {\bm R} = (R_{\sigma 1},\dots, R_{\sigma N})$ and $\sigma{\bm z} = (z_{\sigma 1},\dots, z_{\sigma N})$, respectively.

    The representation \eqref{eq:rw-coupling} of the friction tensors is referred to in \cite{sachs_machine_2025} as a {\em row-wise coupling} ansatz. As detailed in \cite{sachs_machine_2025}, this ansatz is one of several that, together with the stated equivariance properties of the matrix blocks ${\bm \Sigma}_{ij}$, guarantee that the resulting friction tensor ${\bf \Lambda}$ is symmetric positive semi-definite and satisfies the correct equivariance relation with respect to relevant isometric transformations in configurational space, which are
    \[
    \begin{aligned}
    \forall Q \in O(3):& \quad  {\bm \Lambda}_{ij}(Q{\bm R},{\bm z})  = Q{\bm \Lambda}_{ij}({\bm R},{\bm z})Q^{T},\\
    \forall \sigma \in S_N:& \quad {\bm \Lambda}_{ij}(\sigma{\bm R},\sigma{\bm z}) = {\bm \Lambda}_{\sigma i \sigma j}({\bm R},{\bm z}).
    \end{aligned}
    \]

    To expand blocks ${\bf \Sigma}_{ij}$ in an ACE basis, we assume these to be of the functional form
    \[
     {\bm \Sigma}_{ij} = \mathcal{M}^{z_i}_{\rm on}\left (\{(\bm{r}_{ik}, z_{k})\}_{k \in \mathcal{N}(i)}\right ),
     \]
     if $i = j$, and 
     \[
     {\bm \Sigma}_{ij} =\mathcal{M}^{z_i z_j}_{\rm off}\left (\bm{r}_{ij}, \{ (\bm{r}^{\lambda}_{ijk}, z_{k})\}_{k \in \mathcal{N}(i,j)}\right ),
     \]
     for $i \neq j$, where $\bm{r}_{ijk}^{\lambda} := {R}_{k} - (\lambda {R}_i + (1-\lambda) {R}_j)$ for some prescribed $\lambda \in [0,1]$. Here, set notation is used to indicate that the functions $\mathcal{M}^{z_i}_{\rm on}$ and $\mathcal{M}^{z_i z_j}_{\rm off}$ are invariant under permutation of the order of the arguments contained in the respective multisets.
         
     The functions $\mathcal{M}^{z_i}_{\rm on}$ can be linearly expanded as
     \begin{equation}\label{eq:onsite-expansion}
         \mathcal{M}^{z_i}_{\rm on} = \sum_{\eta_{\bm \nu}} c_{z_i,\eta_{\bm \nu}}B^{(2)}_{z_i,\eta_{\nu}},
     \end{equation}
     using a matrix-equivariant ACE basis $(B^{(2)}_{z_i,\eta_{\nu}})_{\eta_{\nu}}$ and coefficients $c_{\eta_{\bm \nu}}\in \mathbb{R}$. This  matrix-equivariant ACE basis is constructed from the atomic basis functions as
        \begin{equation}
    B^{(2)}_{z_i,\eta_{\nu}}=\sum_{\bm{\upsilon}} C^{(2)}_{\eta_{\nu}\bm{\upsilon}} A_{z_i,\bm{\upsilon}}, 
    \end{equation}
    using matrix-valued coupling coefficients $C^{(2)}_{\eta_{\nu}\bm{\upsilon}} \in \mathbb{R}^{3\times 3}$ that are such that each $B^{(2)}_{z_i,\eta_{\nu}}$ satisfies the same equivariance symmetry with respect to $O(3)$ as $ {\bm \Sigma}_{ij}$ (and  $\mathcal{M}^{z_i}_{\rm on}$), i.e.,
    \[
        \begin{aligned}
        \forall Q \in O(3): &\quad  B^{(2)}_{z_i,\eta_{\nu}}\left (\{ (Q \bm{r}_{ik},Z_k) \}_{k \in \mathcal{N}(i)} \right)\\  &= QB^{(2)}_{z_i,\eta_{\nu}}\left (\{ (\bm{r}_{ik},Z_k) \}_{k \in \mathcal{N}(i)} \right) )Q^{T}.
        \end{aligned}
        \]
    An ACE basis expansion of the function $\mathcal{M}^{z_i z_j}_{\rm off}$ can be constructed as a modification of the basis $(B^{(2)}_{z_i,\eta_{\nu}})_{\eta_{\nu}}$; see \cite{sachs_machine_2025,zhang_equivariant_2022} for more details.
    
    To learn a friction tensor from data, a finite subset of ACE basis functions used in the expansion of $\mathcal{M}^{z_i}_{\rm on}$ and $\mathcal{M}^{z_i z_j}_{\rm off}$ is selected and the corresponding coefficients are optimized using as a non-linear optimization procedure that minimizes the discrepancy (measured in terms of a prescribed matrix norm) between evaluations of the resulting parametric model of the friction tensor and evaluations of the EFT obtained from electronic structure calculations.
    
    \subsection{Computational Details} \label{sec:methods_comp_details}
    
        \paragraph{Potential Energy Surface}\label{sec:methods_pes}
    
            We employ MACE-based~\cite{batatia_mace_2022} MLIP models from Ref.~\cite{stark_benchmarking_2024}, trained using a database from Ref.~\cite{stark_machine_2023}. The database included 4,230 structures, with 1,685 Cu-only surface structures sampled at different temperatures (300~K, 600~K, and 900~K), and 2,545 H\textsubscript{2}/Cu structures. The database contains structures including four Cu facets, namely Cu(111), (100), (110), and (211). 3$\times$3, 6-layered slabs for all the surfaces in the database were used with the exception of the Cu(211) surface, for which 1$\times$3, 6-layered slabs were used. The database was built with adaptive sampling, during which structures that are not well represented by the current iteration of the models are selected, labels are generated with DFT, and data points are added to the database. The models are retrained with the updated database. This was repeated 4 times until the final dynamical observable matched the reference values. More details about the database preparation can be found in Ref.~\cite{stark_machine_2023}.
            
            For DFT energy and force calculations, a specific reaction parameter (SRP) functional~\cite{nattino_effect_2012} containing 52\% of PBE~\cite{perdew_generalized_1996} and 48\% of RPBE functional~\cite{hammer_improved_1999} (SRP48) was used with a k grid of 12$\times$12$\times$1 and a tight basis set within the FHI-aims~\cite{blum_ab_2009} all-electron electronic structure code.             Minimum energy paths were obtained using the CI-NEB method, at DFT- and MLIP-level, as detailed in Ref.~\cite{stark_benchmarking_2024}. The two bottom layers of Cu atoms were frozen during optimizations.
                
        \paragraph{LDFA training data generation} \label{sec:methods_train_data_ldfa}

            LDFA models are based on an ML model that predicts the clean metal surface density $\rho_{\mathrm{emb}}(\textbf{R})$ at any given point as a function of the position of the clean surface atoms. This model is used to evaluate the electronic friction coefficient for each H atom independently, from which the LDFA EFT is then constructed. The separate treatment of each H atom is possible due to the isotropic nature of LDFA. We evaluate the the Wigner-Seitz radius $r_{s}(\rho_{\mathrm{emb}})$ 
            \begin{equation}\label{eq:ldfa_wigner}
                r_{s}(\rho_{\mathrm{emb}}) = \left(\frac{3}{4\pi\rho_{\mathrm{emb}}}\right)^{1/3 }
            \end{equation}
            and use an interpolation function (cubic spline) to fit the tabulated LDFA friction coefficients as provided by Gerrits~\textit{et~al.}~\cite{gerrits_electronic_2020}.
        
            To build the density model, a data set is constructed that contains the density values for certain coordinates within the unit cell at different surface configurations. The clean metal surface electronic densities were extracted from densely sampled cube files using the FHI-aims~\cite{blum_ab_2009} electronic structure code, employing ``tight'' settings for basis sets. In total, 200 cube files were generated (50 surface configurations per surface facet).  The cube files were evaluated for surfaces equilibrated at three different temperatures, 300, 600, and 900~K. From each cube file, roughly 30 density values were selected at random positions within the unit cell between the first sublayer and 7~$\mathrm{\AA}$ above the surface. The densities were read at least 0.8~$\mathrm{\AA}$ away from any Cu atom, to avoid high-density peaks close to atom centers that could cause issues in ML model training. The value of 0.8~$\mathrm{\AA}$ was chosen as the minimal distance that can occur between H and Cu atoms at the collision energies of 0.1--1.0~eV. This distance was chosen based on previous scattering simulations. Overall, 6,000 data points were used in the final data set (4,800 in the training data set, and 1,200 in the test set), each containing coordinates of a single H atom and Cu surface atoms, and the corresponding background electronic density of Cu surface atoms at the H atom position. See Fig.~S1 in Supplemental Material for a visualization of the constructed data set for Cu(111).

            The density models were trained using ACEpotentials.jl~\cite{witt_acepotentialsjl_2023} (version 0.6.5). The energies used for the training of ACEpotentials.jl models were replaced with densities, thus, no energies or forces were used in the training of the models. A solver based on Bayesian linear regression (BLR) is used for all the models, within \texttt{ace\_basis} interface. In the final model, a correlation order of 3 was used with the corresponding maximum polynomial degrees of 12, 10, and 10, and with a cutoff distance of 4~$\mathrm{\AA}$. More information about the optimization of model parameters and the final parameters used for the models can be found in Section~\ref{sec:results_valid_ldfa}.
        
        \paragraph{ODF training data generation} \label{sec:methods_train_data_odf}
            The data set for the ODF-based EFT model was constructed by taking a subset of structures from the data set collected for the MACE PES described in our previous publication~\cite{stark_machine_2023}, excluding the structures that contain bare metal surface (1,685 structures), and including only the random subset of H\textsubscript{2}/Cu structures (1,602 out of 2,545). For the Cu(211) surface, a larger slab was required to converge the EFT than was used for the energy and force evaluation (1$\times$4 instead of 1$\times$3). We, therefore, replaced the data points corresponding to this surface (261 structures) with structures generated by sampling new scattering dynamics trajectories (382 structures), which were evaluated using the MACE PES. The sampled trajectories were initialized with different H\textsubscript{2} translational energies, ranging from 0.2~eV to 0.9~eV, and with surface temperatures of 300, 600, and 900~K. The sampling included a clustering procedure using the k-means method. ODF-based EFT was calculated for the prepared set of structures, using Eq.~(\ref{eq:odf_sd}) with a Gaussian broadening width of 0.4~eV as a replacement for the delta function, and a Fermi factor corresponding to an electronic temperature of 300~K. The electronic friction calculations did not require a ``tight'' basis set as needed for PES, and thus ``light'' settings with additional ``second tier'' basis functions for hydrogen (within FHI-aims) were sufficient for the evaluation of ODF-based EFT. Details of the convergence tests are provided in the SM.

            The ODF-based electronic friction models (ACE-friction) were trained using ACEds.jl (\url{https://github.com/ACEsuit/ACEds.jl}, version 0.1.7) code, currently maintained within the ACEfriction.jl package (\url{https://github.com/ACEsuit/ACEfriction.jl})~\cite{sachs_machine_2025}. The ODF calculations were done within the atom-centered scheme (employing column-wise couplings), utilizing L2 regularization with Adam optimizer, for which a learning rate of 10$^{-4}$ was used together with exponential decay rate for the momentum and velocity terms of 0.99, and 0.999, respectively.
            In the final model, a maximum correlation order of 2 was used with the corresponding maximum polynomial degree of 6, and with the cutoff distance of 5~$\mathrm{\AA}$. More information about the optimization of model parameters and the final parameters used for the models can be found in Section~\ref{sec:results_valid_odf}.

        \paragraph{Simulation details} \label{sec:methods_sim_details}
    
            All MD simulations and preparation of initial conditions were performed using the NQCDynamics package~\cite{gardner_nqcdynamicsjl_2022} (\url{https://github.com/NQCD/NQCDynamics.jl}, version 0.13.4).
        
            All the trajectories in this study include H\textsubscript{2} scattering dynamics at different Cu surfaces and are initiated with a hydrogen molecule placed 7~$\textrm{\AA}$ above the surface. Initial conditions for H\textsubscript{2} are established using the Einstein-Brillouin-Keller (EBK) method~\cite{larkoski_numerical_2006} for different normal incidence energies (initial incidence angle of H\textsubscript{2} $\phi=0$\textdegree) and rovibrational states with randomly chosen azimuthal and polar angles. The two bottom metal layers were frozen in all simulations. Simulations at 0~K surface temperatures were initiated using a DFT-relaxed slab with initial velocities based on the Maxwell-Boltzmann distribution. Simulations at finite temperatures were initialized with surface positions obtained from a random sampling of surface-only Langevin MD output structures at the selected temperatures. Lattice expansion is accounted for based on the relation between lattice constant and surface temperature, as described in Ref.~\cite{stark_benchmarking_2024}.
            To ensure that the statistical errors are negligible, sticking, survival, and transition probabilities were calculated using data averaged over 20,000 trajectories for every model, surface facet, rovibrational state, and collision energy reported in this study. More details on statistical errors can be found in the SM. The maximum simulation time of every trajectory was 3~ps with a time step of 0.1~fs unless special conditions were met, such as the distance between hydrogen atoms at the metal surface being above 2.25~$\textrm{\AA}$ or the distance between the hydrogen molecule and top surface atoms being above 7.2~$\textrm{\AA}$. State-to-state survival and transition probabilities are evaluated by dividing the number of trajectories that end with the desired final rovibrational state of H\textsubscript{2} over all the trajectories, including the ones that led to dissociation. The details on the evaluation of dynamical observables described in this study are included in the SM.

\section{Results and Discussion} \label{sec:results}

    \subsection{Validation of Electronic Friction Representations} \label{sec:results_valid}
        In the following sections, we describe nonadiabatic effects within MDEF with two configuration-dependent ML surrogate models of the EFT for LDF and ODF as described in section \ref{sec:methods}. Further details of the data generation and model construction are reported in the SM. Below, we present the details of hyperparameter optimization and validation of models based on evaluation errors, as well as their performance on the dynamical simulations.

        \paragraph*{\textbf{LDFA model}} \label{sec:results_valid_ldfa}

            The optimization (cross-validation) of model parameters was performed for density models based on ACEpotentials.jl code. In Fig.~S2 (SM), the model size (correlation order and corresponding polynomial degrees) and cutoff distance are plotted against the root mean squared errors (RMSEs). Additionally, density evaluation times obtained for the corresponding models are included. ACE-based density models achieve rapid convergence with respect to both model size and cutoff distance, resulting in low evaluation times of roughly 15~ms (per single H at the Cu surface). Even considering the separate evaluations for each H atom, this is almost two times faster than the force evaluation with our MACE PES model (roughly 50~ms). The model RMSE converges already for correlation order (N) of 2 with higher polynomial degrees (10 and 6 or 16 and 2). Full convergence is already achieved with a cutoff distance of 3~$\mathrm{\AA}$. In the final density model, the correlation order of 3 is used with the corresponding polynomial degrees (for each order) of 12, 10, and 8, respectively. In the final model, a cutoff distance of 4~$\mathrm{\AA}$ was used. 

            Aside from several outliers corresponding to the Cu(111) surface, the ACE predicted density values match the reference densities across the entire spectrum of the densities in the test set as shown in Fig.~S3 (SM). The maximum difference from the reference is 0.05~$\mathrm{\AA}^{-3}$. The final RMSE and mean absolute error (MAE) are 0.004~$\mathrm{\AA}^{-3}$ and 0.002~$\mathrm{\AA}^{-3}$.
    
            \begin{figure}
                \centering
                \includegraphics[width=1.0\linewidth]{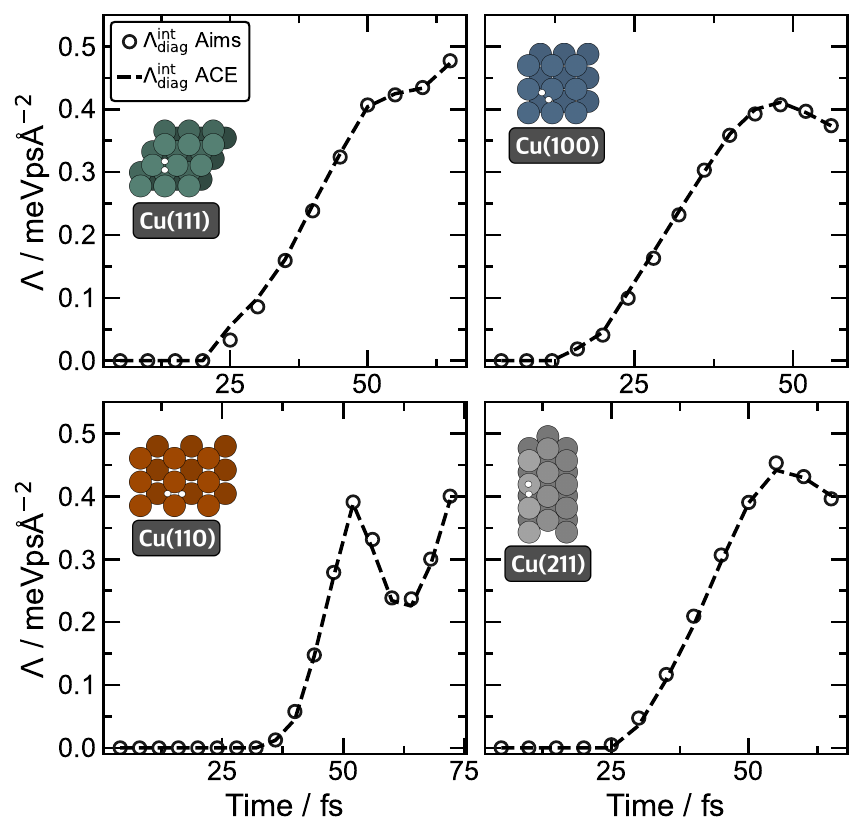}
                \caption[Predictions of LDFA-based EFT elements along the trajectories of H\textsubscript{2} dissociation on Cu surfaces.]{\textbf{Predictions of LDFA-based EFT elements along the trajectories of H\textsubscript{2} dissociation on Cu surfaces.} The diagonal elements of EFT (internal coordinates) are plotted for dissociation trajectories at Cu surfaces, predicted by the ACE model, and compared to reference values calculated using FHI-aims. Note that all the diagonal components of the EFT are identical in normalized internal coordinates. Criteria to identify dissociation are given in section \ref{sec:methods}.}
                \label{fig:ldfa_model_val_dissoc}
            \end{figure}
            
       The final predicted LDFA values show excellent agreement with DFT values along simulated dissociation trajectories (Fig.~\ref{fig:ldfa_model_val_dissoc}) for the four different copper surfaces, namely, Cu(111), Cu(100), Cu(110), and Cu(211). The shown trajectories were chosen randomly from the trajectories run with MACE PES. The LDFA-based electronic friction along the diagonal elements generally increases as the molecule approaches the surface across all considered copper facets. Notably, for Cu(110), the friction exhibits a distinct drop just before dissociation, whereas for Cu(211) and Cu(100), the friction peaks prior to dissociation. After dissociation, the electronic friction reaches its highest values for hydrogen atoms on the Cu(111) surface. The friction for dissociated hydrogen atoms is approximately 0.4~meVpsÅ\textsuperscript{-2} for Cu(100), Cu(110) and Cu(211), while for Cu(111), it is slightly elevated at around 0.47~ meVpsÅ\textsuperscript{-2}. A similar comparison is shown in Fig.~S4 (Supplemental Material) for scattering trajectories, where the excellent agreement between LDFA based on ACE and DFT holds across all facets. 
            
        \paragraph*{\textbf{ODF} model} \label{sec:results_valid_odf}

            The convergence of the DFT calculations of the ODF EFT has been carefully studied. We find that the dd element convergence is slower than for the ZZ element with respect to k-grid size as shown (Fig.~S5, SM). 
            However, the convergence of all the considered surfaces is consistent, and a satisfactory level of convergence ($\pm$10\% variation in the EFT values) is reached around N\textsubscript{k}=12 for both investigated elements. 
            A Gaussian broadening width of 0.4~eV was chosen to evaluate ODF, as this is a compromise between achieving stable EFT prediction and excessive broadening across the surface facets (Fig.~S6, SM).
            In most previous studies, the broadening widths of 0.6 were chosen.~\cite{spiering_testing_2018,box_determining_2021} Large broadening widths can lead to the incorporation of non-zero frequency contributions that are not included in the first-order approximation within many-body perturbation theory. Our results are relatively insensitive to the choice of broadening within the range of 0.3--0.8~eV. A `light' basis is employed, as defined in FHI-aims, as ODF coefficients were already converged with this choice of basis set.

            A $3\times$3 unit cell was chosen for the calculations involving Cu(111), Cu(110), and Cu(100) surfaces, and 1$\times$4 unit cell for the Cu(211) surface, with 6 layers for all surfaces, as these gave stable ODF coefficient predictions (Fig.~S7, SM).

            A cross-validation optimization of the ODF ACE-friction model parameters was performed, in which an 80\%/20\% train/test split was used. Different settings were tested for maximum correlation order and maximum polynomial degree (Fig.~S8, SM).

            ODF models were trained with maximum correlation orders from 1 to 3 (left side of Fig.~S8). RMSEs improved up to order 2, with no further gains at higher orders. Since higher orders significantly increase evaluation time, order 2 was chosen for the final models.
            ODF models were trained with maximum polynomial degrees from 2 to 10 for a correlation order of 2 (right side of Fig.~S8). A degree of 6 was found to be optimal, with higher degrees reducing training RMSEs but increasing test RMSEs, likely due to overfitting. The EFT evaluation time for the ODF ACE-friction model is ~40 ms, only slightly slower than the LDFA-based EFT (~30 ms), but it computes the full EFT rather than just two diagonal values. Additionally, RMSEs were tested for cutoff distances from 3 to 7~$\mathrm{\AA}$ (Fig.~S9, SM). RMSEs improve rapidly between 3 and 4~$\mathrm{\AA}$, with only slight improvement up to 5~$\mathrm{\AA}$. Beyond this, RMSEs increase, likely due to cutoff overlap between periodic unit cells. A 5~$\mathrm{\AA}$ cutoff was chosen for the final models. The model performance is shown in Fig.~S10 (SM), comparing ACE-friction predictions with DFT reference values. The predictions align well with the reference data (MAE of 0.0038~meVps$\mathrm{\AA}^{-2}$ and RMSE of 0.0083~meVps$\mathrm{\AA}^{-2}$), with some outliers for higher electronic friction values.

            \begin{figure}
                \centering
                \includegraphics[width=1.0\linewidth]{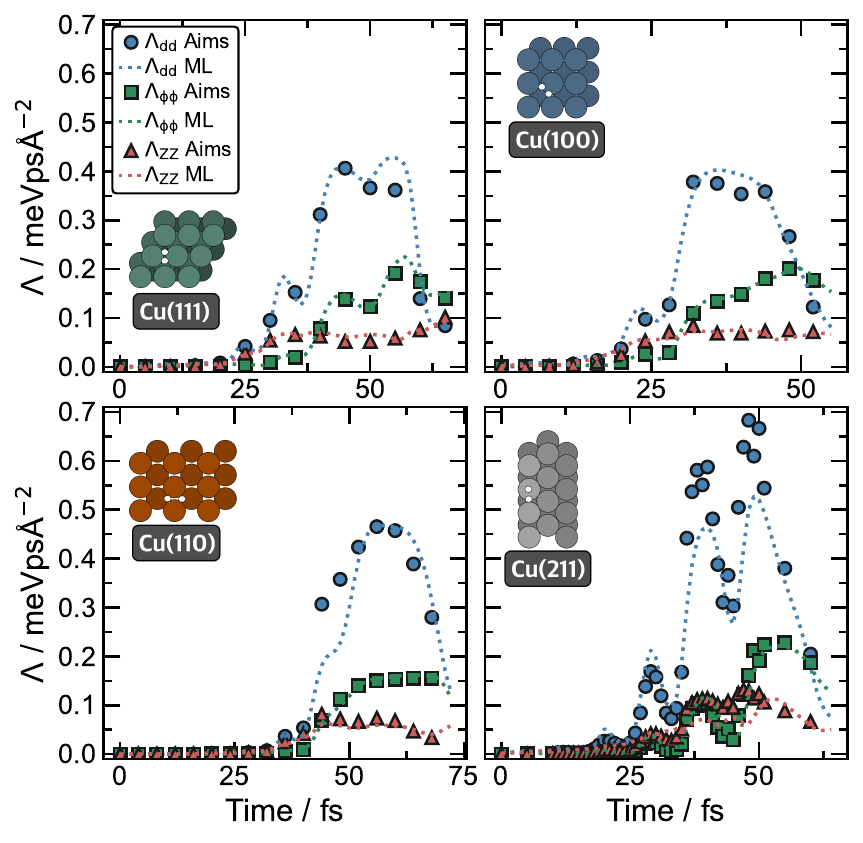}
                \caption{\textbf{Predictions of ODF-based EFT elements across trajectory steps of H\textsubscript{2} dissociation on Cu surfaces.} Electronic friction values (internal coordinates) are plotted for dissociation trajectories (the same as in Fig.~\ref{fig:ldfa_model_val_dissoc}) at Cu surfaces predicted by the ACE-friction model and compared to reference values calculated using FHI-aims. Dissociation criteria are given in the section \ref{sec:methods}.}
                \label{fig:odf_model_val_dissoc}
            \end{figure}

            To assess the accuracy of the final ACE-friction model, EFT was evaluated along scattering (Fig.~S11, SM) and dissociation (Fig.~\ref{fig:odf_model_val_dissoc}) trajectories using both the ACE-friction model and the reference DFT data. This evaluation covered all four Cu surfaces in this study. Three electronic friction components in the internal coordinate system were compared: internal stretch of the adsorbate atoms (d), azimuthal angle ($\mathrm{\phi}$), and the center of mass in the z direction (Z) (see SM for details on the coordinate transformation). The ACE-friction model predictions generally agree well with the reference values (Aims). However, notable discrepancies appear in some cases, such as $\Lambda_{\mathrm{\phi\phi}}$ in H\textsubscript{2} scattering on Cu(110) and $\Lambda_{\mathrm{dd}}$ in H\textsubscript{2} dissociation on Cu(211). This behavior persists in models retrained with different settings (\textit{e.g.}, with a maximum correlation order of 3). Similar to the LDFA case, the electronic friction increases on approach to the surface, however, the final dissociation product states exhibit significantly lower electronic friction values (maximum of 0.15~meVps$\mathrm{\AA}^{-2}$) than the corresponding LDFA values, which exceed 0.35~meVps$\mathrm{\AA}^{-2}$ across all surfaces.
            In some cases, particularly near the dissociation barrier where the EFT values peak, the ACE-friction model deviates more significantly from the reference data. These deviations likely do not translate into large errors in friction forces as they occur in the transition state regions where velocities are low (see Figs.~S16 and S17, SM). As per eq. \ref{eq:2_mdef}, the corresponding effect on the friction forces will be small.

        \paragraph*{\textbf{Electronic friction along minimum energy paths}} \label{sec:results_valid_meps}
        
        \begin{figure}
            \centering
            \includegraphics[width=\linewidth]{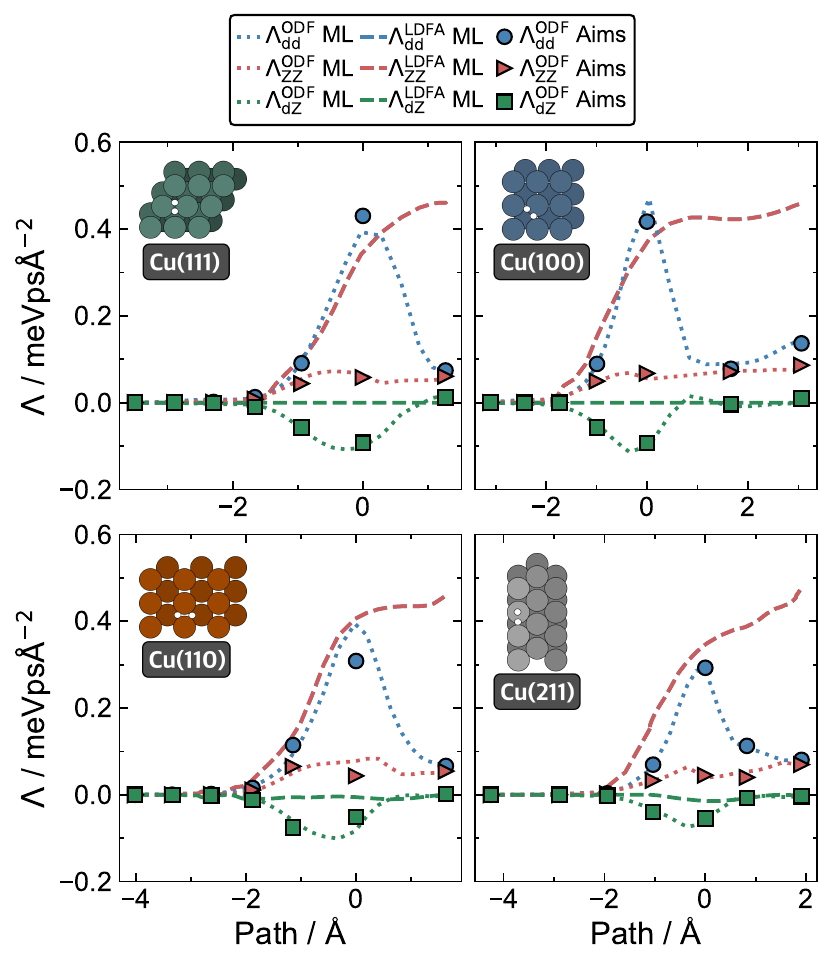}
            \caption{\textbf{Electronic friction values along minimum energy paths of dissociative adsorption of H\textsubscript{2} at Cu surfaces.} EFT elements ($\Lambda_{\mathrm{dd}}$, $\Lambda_{\mathrm{dZ}}$, and $\Lambda_{\mathrm{ZZ}}$) are shown for different Cu surfaces along the minimum energy paths, evaluated with FHI-aims (ODF), and two ML-based models, based on LDFA (dashed lines) and ODF methods (solid lines). The transition state along the path in each case is located at 0~$\mathrm{\AA}$.  Friction values are shown in the internal coordinate system (Fig.~\ref{fig:int_cart_coord}), where d corresponds to an interatomic distance between H atoms, and $\theta$ and $\phi$ to polar and azimuthal angles with respect to the surface normal. X, Y, and Z represent the center-of-mass positions of the adsorbate molecule.}
            \label{fig:mep_neb_eft}
        \end{figure}

             We evaluated electronic friction along the minimum energy paths for H\textsubscript{2} dissociative chemisorption on Cu(111), Cu(100), Cu(110), and Cu(211) surfaces using reference FHI-aims ODF calculations, as well as our ODF and LDFA models (Fig.~\ref{fig:mep_neb_eft}). The ODF ML model closely follows the reference electronic friction values, which for all surface facets exhibit an increase in EFT elements to the transition state, and then a decrease to the dissociated product state with few exceptions, such as the $ZZ$ element for Cu(211). The LDFA model produces isotropic friction values, resulting in identical internal EFT elements when using a unitary coordinate transformation. While LDFA values around the transition state align with the $\Lambda_{\mathrm{dd}}$ ODF values (approximately 0.4~meVps$\mathrm{\AA}^{-2}$), they increase further for the final dissociation structure. This contrasts with the ODF values, which decrease to near 0.1~meVps$\mathrm{\AA}^{-2}$ in the final structure.
             As discussed in Ref~\cite{box_room_2024} for H/Pt(111), LDFA significantly overestimates electronic friction for chemisorbed hydrogen atoms on metal surfaces, resulting in much shorter vibrational lifetimes than observed experimentally. The discrepancy between LDFA and ODF for the chemisorbed state is consistent across all studied copper facets. Previous ODF results~\cite{spiering_testing_2018, luntz_comment_2009,luntz_how_2005} for H\textsubscript{2}/Cu(111) used $\Lambda^{0}$, which showed a smaller disparity between LDFA and ODF in the product state.~\cite{box_ab_2023} Fig.~S13 in the SM shows a comparison of our EFT values along H\textsubscript{2}/Cu(111) dissociation minimum energy path with previous works~\cite{spiering_testing_2018, luntz_comment_2009,luntz_how_2005}. 

    \subsection{Validation Against Experiment}

          \begin{figure}
            \centering
            \includegraphics[width=1.0\linewidth]{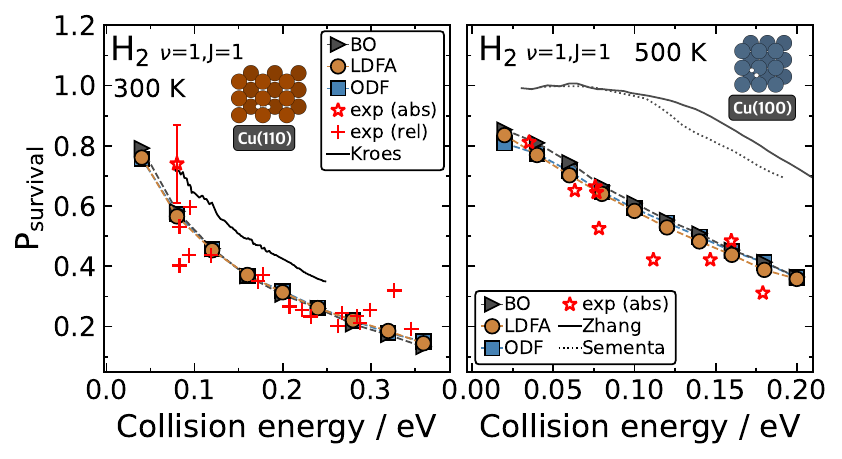}
            \caption{\textbf{Survival probabilities for the state-to-state scattering of H\textsubscript{2} at Cu(110) and Cu(100) surfaces.} Survival probabilities of H\textsubscript{2} ($\mathrm{\nu}$=1, J=1) at Cu(110) and Cu(100) surfaces are plotted for a set of translational energies (in eV) (at 300~K and 500~K, respectively) and compared with previous computational and experimental results. The probabilities were calculated using classical BOMD (triangles), MDEF-LDFA (circles), and MDEF-ODF (squares). Survival probabilities for H\textsubscript{2} scattering at Cu(110) are compared with theoretical results by Kroes~\textit{et~al.}~\cite{kroes_dissociative_2007} (solid black line) and relative experimental survival probabilities by Gostein~\textit{et~al.}~\cite{gostein_scattering_1996} scaled such that the highest collision energy matches our simulations (red crosses). A single available absolute experimental probability by Gostein~\textit{et~al.}~\cite{gostein_survival_1995} is also included with a corresponding error bar (red star). For H\textsubscript{2} scattering at Cu(100), the survival probabilities are compared with absolute (non-scaled) experimental survival probabilities by Watts~\textit{et~al.}~\cite{watts_state--state_2001} (red stars), and theoretical results by Zhang~\textit{et~al.}~\cite{zhang_six-dimensional_2022} (solid black line) and Sementa~\textit{et~al.}~\cite{sementa_reactive_2013} (dotted black line).
            }
            \label{fig:results_sts_survival_exp}
        \end{figure}
        
      We now turn to benchmarking our simulations against experimental measurements and prior theoretical studies. These comparisons allow us to critically assess the accuracy and transferability of the MLIP and EFT ML models across observables and to evaluate the impact of electronic friction on inelastic energy redistribution during state-to-state scattering. While recent work has shown that adiabatic BOMD simulations based on our MLIP yield good agreement with experimental sticking probabilities on Cu(111) and Cu(211)~\cite{stark_benchmarking_2024}, we now focus on the quantum state distribution of scattered molecules where the role of nonadiabatic effects may be more significant.
        
        Rovibrational state survival probabilities for H\textsubscript{2} ($\mathrm{\nu}$=1, J=1) scattering on Cu(110) at 300~K were compared with experimental results (single absolute survival probability data point and set of relative probabilities) by Gostein~\textit{et~al.}~\cite{gostein_survival_1995,gostein_scattering_1996} and theoretical predictions by Kroes~\textit{et~al.}~\cite{kroes_dissociative_2007}. The latter results are based on time-dependent quantum wavepacket dynamics based on the PW91 functional and a static surface approximation (Fig.~\ref{fig:results_sts_survival_exp}, left).

        All methods and experiments show a decrease in survival probabilities with increasing collision energy.
        Absolute survival probabilities by Kroes~\textit{et~al.}~\cite{kroes_dissociative_2007} are consistently about 0.1 larger than our BOMD, MDEF-LDFA, and MDEF-ODF results across all energies. However, the results of Kroes~\textit{et~al.} capture the relative decrease in survival probability as well as our results. 
        The absolute offset in survival probability may stem from their use of the PW91 functional, whereas we employed SRP48. 
        Another factor could be our inclusion of surface movement at 300~K, in contrast to the rigid surface model used by Kroes~\textit{et~al.}~\cite{kroes_dissociative_2007}. 
        The inclusion of electronic friction shows a negligible effect when compared to BOMD, suggesting that nonadiabatic effects have little impact on the rovibrational survival probability of H\textsubscript{2} ($\mathrm{\nu}$=1,J=1) in scattering at Cu(110).
        Only a single absolute survival probability is available from experiments, and due to a relatively large error (0.13), we cannot make a firm judgment on which theoretical curve gives a more accurate prediction.

        All our models agree well with absolute survival probabilities for H\textsubscript{2}($\mathrm{\nu}$=1,~J=1) elastic scattering at Cu(100) surface at 500~K, measured by Watts~\textit{et~al.}~\cite{watts_state--state_2001}, across all studied collision energies (Fig.~\ref{fig:results_sts_survival_exp}, right). 
        This is a significant improvement over  previous results by Sementa~\textit{et~al.}~\cite{sementa_reactive_2013} using an adiabatic approach, employing a PES based on the SRP43 functional (43\% RPBE, 57\% PW91) and Zhang~\textit{et~al.}~\cite{zhang_six-dimensional_2022} using a quantum-dynamical approach employing a time-dependent wave packet method with a different transferable MLIP~\cite{zhu_unified_2020} based on the optPBE-vdW functional~\cite{klimes_chemical_2010}. Both the reference theoretical results use rigid metal surfaces in their simulations and made similar predictions of survival probabilities, however, both significantly overestimated the survival probabilities compared to the experiment. Such improvement in accuracy could be attributed to either the inclusion of surface movement in the dynamics, which will have a significant effect at the experimental temperature of 500~K, or by the difference in exchange-correlation functional used. The negligible difference with or without friction again suggests that nonadiabatic effects are not significant for survival probabilities. However, an important benefit of the combined MLIP and EFT ML models is that we do not need to make a prior judgment whether to include nonadiabatic effects in our simulations or not.

    \begin{figure}[h!]
        \centering
        \includegraphics[width=0.75\linewidth]{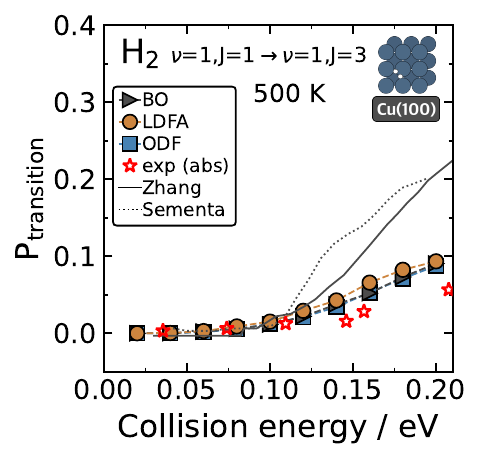}
        \caption{\textbf{Rotational excitation probabilities for the state-to-state scattering of H\textsubscript{2} at Cu(100) surface (500~K).} The H\textsubscript{2} ($\mathrm{\nu}$=1,~J=1~$\rightarrow$~$\mathrm{\nu}$=1,~J=3) probabilities were evaluated at different collision energies employing three different schemes, adiabatic (BOMD) (black triangles), MDEF with LDFA (orange circles), and MDEF with ODF (blue squares), using MACE-based PES, ACE density model (LDFA), and ACE-friction model (ODF). The H\textsubscript{2} ($\mathrm{\nu}$=1,~J=1~$\rightarrow$~$\mathrm{\nu}$=1,~J=3) transition probabilities are also compared with absolute (non-scaled) experimental survival probabilities by Watts~\textit{et~al.}~\cite{watts_state--state_2001} (red stars) and theoretical results by Zhang~\textit{et~al.}~\cite{zhang_six-dimensional_2022} (solid black line) and Sementa~\textit{et~al.}~\cite{sementa_reactive_2013} (dotted black line).}
        \label{fig:h2cu_rot_exc_j1_j3_cu100_exp}
    \end{figure}

        In the same experimental work by Watts~\textit{et~al.}~\cite{watts_state--state_2001}, experimental H\textsubscript{2}($\mathrm{\nu}$=1,~J=1~$\rightarrow$~$\mathrm{\nu}$=1,~J=3) absolute rotational excitation probabilities were collected for scattering dynamics at Cu(100) at 500~K. 
        In Fig.~\ref{fig:h2cu_rot_exc_j1_j3_cu100_exp}, simulated transition probabilities for rotational excitation of H\textsubscript{2} ($\mathrm{\nu}$=1,~J=1~$\rightarrow$~$\mathrm{\nu}$=1,~J=3) at Cu(100) at 500~K are compared with experimental results by Watts~\textit{et~al.}~\cite{watts_state--state_2001} and theoretical results by Zhang~\textit{et~al.}~\cite{zhang_six-dimensional_2022} and Sementa~\textit{et~al.}~\cite{sementa_reactive_2013}. Close agreement with experimental results is obtained with the BOMD method, and no further improvement is visible when nonadiabatic effects are considered in the dynamics (using MDEF with either LDFA or ODF EFT representations). Such close agreement was not achieved by previous theoretical results that deviated from experiment for collision energies above 0.1~eV. At the collision energy of 0.2~eV, the previous theoretical works predict transition probabilities of about 0.22, overestimating the probabilities by a factor of four compared to the experimental results. 

        Based on our validation results, we can conclude that the combination of the MACE MLIP developed by Stark~\textit{et al.}\cite{stark_benchmarking_2024} and the EFT ML models presented here, represent the to date most accurate models to simulate reactive surface chemistry of hydrogen on low-index facets of copper. 

    \subsection{Dissociative Chemisorption and Rovibrationally Elastic Scattering Across Surfaces} \label{sec:results_sticking}

\begin{figure}
            \centering
            \includegraphics[width=1.0\linewidth]{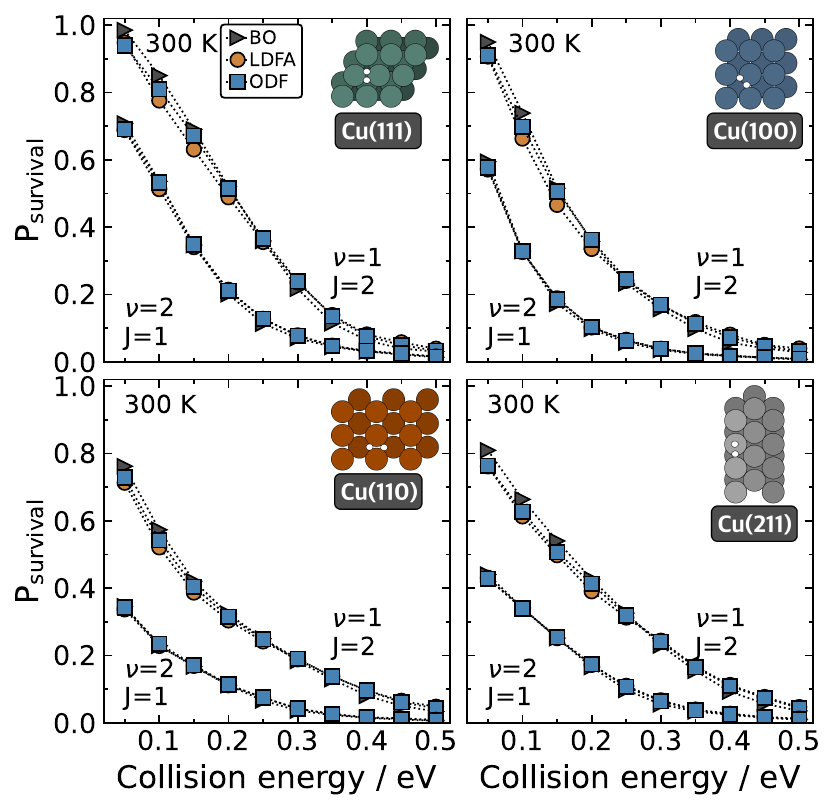}
            \caption{\textbf{Survival probabilities for the state-to-state scattering of H\textsubscript{2} at Cu surfaces.} Survival probabilities of H\textsubscript{2} ($\mathrm{\nu}$=1, J=2 and $\mathrm{\nu}$=2, J=1) at Cu surfaces are plotted for a set of translational energies (eV) at 300~K. The probabilities were calculated using classical BOMD (triangles), MDEF-LDFA (circles), and MDEF-ODF (squares).}
            \label{fig:results_sts_survival}
        \end{figure}

\begin{figure}
            \centering
            \includegraphics[width=1.0\linewidth]{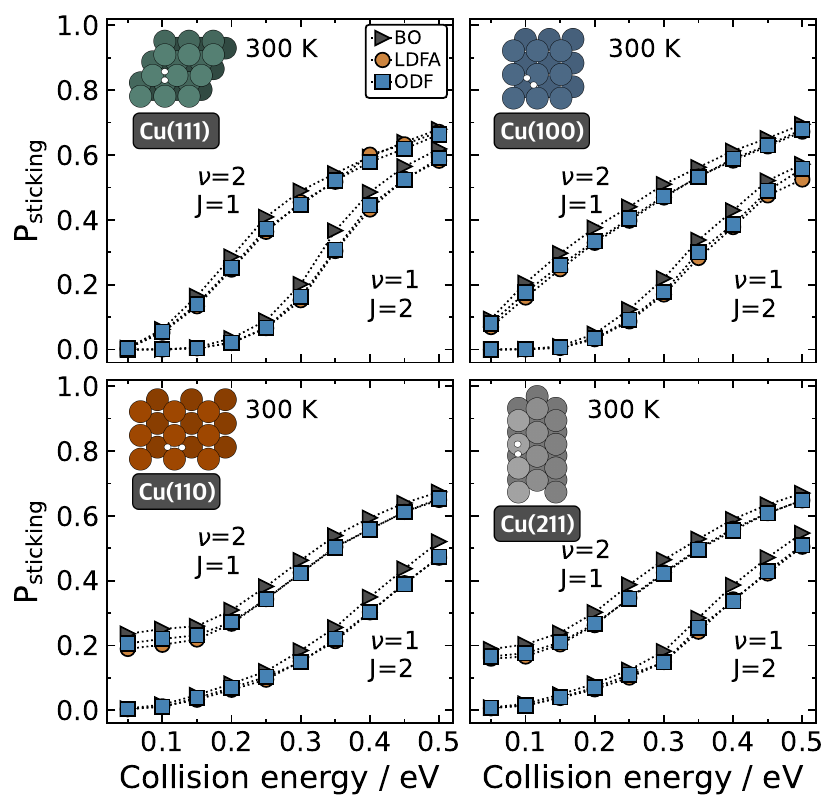}
            \caption{\textbf{Sticking probabilities evaluated at 300~K for H\textsubscript{2} dissociation on different Cu surfaces.} The probabilities were evaluated at different collision energies employing three different schemes, adiabatic (BO), MDEF-LDFA, and MDEF-ODF, using MACE-based PES, ACE density model (LDFA), and ACE-friction model (ODF). The results were obtained for two H\textsubscript{2} rovibrational initial states, ($\bm{\nu}$=2, J=1) and ($\bm{\nu}$=1, J=2).}
            \label{fig:h2cu_sticking}
        \end{figure}

        Having established that the model is capable of simulating several phenomena correctly across different surface facets, we investigate how the surface morphology influences H\textsubscript{2} dissociation and scattering. To do so, we evaluate sticking and survival probabilities across four low-index Cu surfaces—Cu(111), Cu(100), Cu(110), and Cu(211)—using both adiabatic and nonadiabatic MD simulations  (Figs.~\ref{fig:results_sts_survival} and ~\ref{fig:h2cu_sticking}). Two initial rovibrational states were considered: ($\nu$=1, J=2) and ($\nu$=2, J=1). For all surfaces and methods, survival probabilities monotonically decrease with increasing collision energy, while sticking probabilities increase, consistent with barrier-limited dynamics.

        The inclusion of electronic friction (LDFA and ODF) introduces only minor changes across the full energy range, indicating that these processes are predominantly adiabatic under the conditions studied. Differences between friction models are negligible, and trends in sticking and survival probabilities are robust against the choice of the method of evaluating the EFT.

        Across all investigated surfaces, H\textsubscript{2} in the $\nu=1$ vibrational state exhibits negligible sticking at the lowest collision energies. As energy increases, sticking rises first on Cu(110) and Cu(211), reflecting their greater reactivity at low incidence energies, while Cu(100) and Cu(111) show a delayed but sharper increase at higher energies. When the molecule is initially in the $\nu=2$, J=1 state, vibrational excitation significantly enhances reactivity on Cu(110) and Cu(211), resulting in appreciable sticking even at low translational energies. In contrast, Cu(100) and Cu(111) remain relatively inert at these energies, with sticking rising more steeply only at higher collision energies. These trends highlight the role of vibrational promotion, particularly on Cu(110) and Cu(211) surfaces.
        
        Experimental measurements by Anger et al.~\cite{anger_adsorption_1989} using molecular beam techniques report initial sticking coefficients for H\textsubscript{2} on Cu(111), Cu(100), and Cu(110) across a similar energy range. While these measurements do not resolve rovibrational states, they indicate a sticking order of Cu(111) $>$ Cu(100) $>$ Cu(110), which contrasts with our results for $\nu=2,\ J=1$ but more closely resembles our $\nu=1,\ J=2$ results. This suggests that the experimental measurements may reflect behavior closer to the vibrational ground state ($\nu=0$), where vibrational promotion is absent or less pronounced.

        \begin{table}[h!]
           \centering
           \caption{\textbf{Reaction barriers, $E_\text{ads}$, associated with dissociative chemisorption of H\textsubscript{2} at different Cu surfaces.} The reaction barriers obtained for H\textsubscript{2} dissociation at Cu(111), Cu(100), Cu(110), and Cu(211) are listed, based on the CI-NEB calculations. Corresponding values from literature{---}labeled as $E_\text{ads}^\text{lit}${---}with the closest settings to our calculations are given for comparison's sake, too. Additionally, vibrational promotion energies ($\Delta E_{\mathrm{\nu}=2\rightarrow\mathrm{\nu}=1}$) are listed, which describe the difference in energy between the same sticking probability (0.3) at two different vibrational states $\mathrm{\nu}=1$ and $\mathrm{\nu}=2$. The DFT calculations were performed with the FHI-aims code, employing the SRP48 functional.}
           \begin{tabular}{l|llll} \hline 
               Surface & Cu(111) & Cu(100) & Cu(110) & Cu(211) \\ \hline
               $E_\text{ads}$(this work) / eV & 0.56 & 0.58  & 0.63  & 0.52  \\ 
               $E_\text{ads}$(lit) / eV & 0.64\footnote{xc=SRP48; $p(3\times3)$ cell; Static surface{--}taken from Ref.\cite{smeets_designing_2021}}  & 0.67\footnote{\label{note2}xc=optPBE-vdW; $p(2\times2)$ cell; Moving surface{--}taken from Ref.\cite{zhu_unified_2020}} & 0.81$^\text{b}$ & 0.64\footnote{xc=SRP48; $p(3\times3)$ cell; Static surface{--}taken from Ref.\cite{cao_hydrogen_2018}}\\ \hline
                $\Delta E_{\mathrm{\nu}=2\rightarrow\mathrm{\nu}=1}$ / eV & 0.13 & 0.17 & 0.18 & 0.15 \\ \hline 
           \end{tabular}
           \label{tab:barriers}
        \end{table}

        To rationalize these trends, we examine the computed dissociation barriers $E_\text{ads}$ and vibrational promotion energies $\Delta E_{\nu=2\rightarrow\nu=1}$ (Table~\ref{tab:barriers}). Cu(110) exhibits the highest barrier (0.63 eV) and the strongest vibrational enhancement (0.18 eV), followed by Cu(100). Cu(111) has the next highest barrier (though not necessarily distinguishable within the computational uncertainty), but the weakest vibrational promotion. Despite its high barrier, Cu(110) shows high reactivity at low collision energy for $\nu=2$, suggesting that geometrical effects and vibrational promotion facilitate access to the transition state.

        These observations are consistent with the shape of the elbow plots (2D slices of the PES around the transition state): Cu(111) presents an early barrier, while Cu(110) and Cu(100) exhibit mid-barrier character, and Cu(211) shows a later transition state that is where vibrational promotion will be more beneficial (Fig.~S14, SM). 
        This aligns with Polanyi-type behavior, where the location of the transition state dictates the effectiveness of vibrational excitation in promoting dissociation. For example, Cu(111) shows a stronger increase in sticking in response to increasing collision energy than Cu(110) and Cu(211).
        
    \subsection{Rovibrationally Inelastic Scattering across Surfaces} \label{sec:results_inelastic_scattering}

\begin{figure}
            \centering
            \includegraphics[width=1.0\linewidth]
            {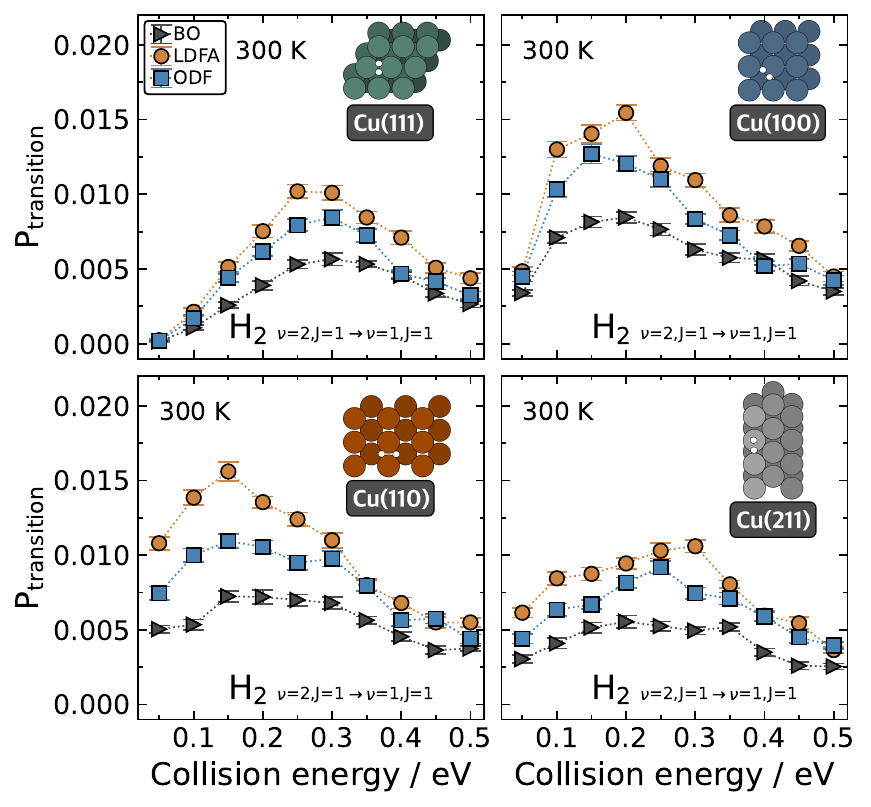}
            \caption{\textbf{Vibrational de-excitation probabilities for the state-to-state scattering of H\textsubscript{2} at Cu surfaces (300~K).} Vibrational de-excitation of H\textsubscript{2}($\mathrm{\nu}$=2,~J=1~$\rightarrow$~$\mathrm{\nu}$=1,~J=1) at Cu(111), Cu(100), Cu(110), and Cu(211) surfaces are plotted for a set of translational energies (eV) at 300~K. The probabilities were calculated using classical BOMD (triangles), MDEF-LDFA (circles), and MDEF-ODF (squares). Statistical error bars shown, see SM for details.
            }
            \label{fig:results_sts_v2J1_vibr}
        \end{figure}

        We next examined vibrational energy loss in rovibrationally inelastic scattering (Fig.~\ref{fig:results_sts_v2J1_vibr}), focusing on the transition $\nu=2,\ J=1$ $\rightarrow$ $\nu=1,\ J=1$ at $300$~K across all Cu surfaces. Vibrationally inelastic transition probabilities are low and weakly dependent on the incidence energy under adiabatic dynamics, but increase with a shift in peak position on the inclusion of nonadiabatic effects.
  
        For all surfaces, nonadiabatic dynamics (LDFA and ODF) yield a higher dependence on the collision energy than adiabatic dynamics. The two schemes do not differ strongly from each other except in the case of the Cu(110), where LDFA predicts higher vibrational de-excitation probabilities at low energies. Whilst scattering at Cu(110) and Cu(100) produce pronounced peaks in transition probabilities at low collision energies ($\approx0.15$~eV) with LDFA, scattering at Cu(211) and Cu(111) show a broader, flatter peak in transition probability at intermediate energies (0.25–0.3 eV), reflecting the complicated geometric effects at play. At high energies ($>$0.4 eV), transition probabilities decline across all surfaces. This is consistent with the rise in dissociative sticking, which removes trajectories from the scattering channel. In the case of Cu(100) and Cu(110), the collision energy dependence that arises from nonadiabatic effects may indeed be measurable in the future.

        We further compare rovibrational de-excitation probabilities for H\textsubscript{2} and D\textsubscript{2} on Cu(111) at 0~K with prior theoretical work by Spiering and Meyer~\cite{spiering_testing_2018} Fig.~S15, SM). 
        While our adiabatic and LDFA results are in good agreement, we find much closer agreement between ODF and LDFA than previously reported. This may indicate that the previously reported fingerprint of specifically anisotropic nonadiabatic effects reported for Cu(111) may be an artefact of the low-temperature approximation (eq.~\ref{eq:odf_dd}) employed for ODF.~\cite{calandra_electron-phonon_2005,box_ab_2023}. This finding would also be consistent with previous MDEF(ODF) simulations for H$_2$ scattering on Ag(111).~\cite{zhang_hot-electron_2019, maurer_hot_2019} Our results suggest that the largest difference between ODF and LDFA can be found for Cu(110) at low collision energies. Comparison between the 0~K and 300~K simulations further suggests that surface phonons play only a marginal role for scattering at Cu(111).

      Taken together, these results show that while rovibrational de-excitation is a finely resolved and potentially sensitive observable, owing to the population being distributed over many rotational states, the absolute impact of electronic friction remains small across all surfaces studied and experimental resolution would have to significantly increase to be able to resolve subtleties that may arise from nonadiabatic effects.

\subsection{Vibrational Transition Landscapes}

\begin{figure}
            \centering
            \includegraphics[width=1.0\linewidth]{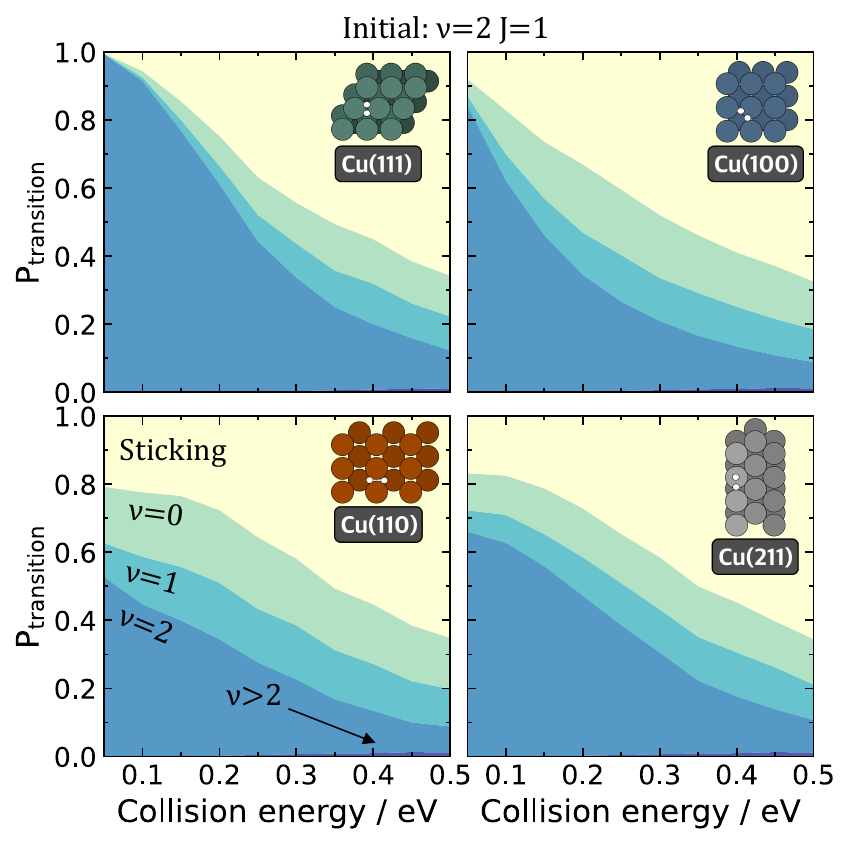}
            \caption{\textbf{
            Vibrational transition probabilities as stacked area plots for H\textsubscript{2} scattering at Cu surfaces (300~K).} Final states after scattering of H\textsubscript{2}($\mathrm{\nu}$=2,~J=1) are labeled. All final rotational quantum numbers are included. The vibrational excited population ($\mathrm{\nu}>2$) is very small. The probabilities were calculated using MDEF-ODF.
            }
            \label{fig:odf_contour_v2}
        \end{figure}

    To provide a comprehensive picture of the scattering dynamics across different Cu facets, we present an overview of the final state distributions in Fig.~\ref{fig:odf_contour_v2}. These vibrational ``transition landscapes'' visualize the fate of H\textsubscript{2} molecules initially prepared in the rovibrational state $\nu=2$, J=1 as a function of collision energy at 300~K. Using stacked area plots from MDEF-ODF simulations, we distinguish between vibrationally (in)elastic scattering, as well as dissociation outcomes.
    
    Each surface exhibits a characteristic landscape reflecting its underlying reactivity and vibrational mode-specific energy transfer behavior. For Cu(111), most molecules remain in vibrationally excited bound states at low energies, with dissociation becoming dominant only above $\sim$0.3--0.4~eV. In contrast, Cu(110) and Cu(211) show broader and more gradual transitions: vibrational de-excitation is more prominent at low energies, and dissociation occurs even at the lowest energies, consistent with their enhanced low-energy reactivity. Cu(100) displays low energy reactivity and generally intermediate behavior between the response of Cu(111) and the broader distributions seen for Cu(110) and Cu(211).
    
    The fraction of molecules that remain in the initial state decreases with increasing energy across all surfaces, as population flows into lower vibrational states and eventually into the dissociated channel. The population of molecules undergoing vibrational excitation ($\nu>2$) is negligible, confirming that vibrational excitation is rare under these thermal conditions.
    
    We also evaluated the equivalent transition landscapes using adiabatic and MDEF(LDFA) dynamics (Sec.~XI, SM) and found only subtle quantitative differences. This further supports the conclusion that the role of electronic friction is subtle in this system and does not dramatically reshape the outcomes of reactive scattering for H\textsubscript{2} on copper.
    
    Together, these transition landscapes provide an intuitive and surface-specific summary of how energy and population flow through the scattering event. They highlight the influence of surface structure on dissociation thresholds, vibrational quenching, and the balance between elastic and inelastic channels, offering a concise visual representation of the full dynamical fingerprint for each facet.
                
\section{Conclusions and outlook} \label{sec:conclusions}

In this work, we present machine learning surrogate models of electronic friction tensors that are transferable across different surface facets of copper. Both models are accurate representations of the underlying first principles data. In combination with previously published machine learning interatomic potentials, they constitute a framework to simulate quantum-state-resolved dynamics of H\textsubscript{2} scattering and dissociation on a range of Cu surfaces with high accuracy and with the ability to perform comprehensive statistical sampling. These models accurately reproduce reference data along representative scattering and dissociation pathways, enabling efficient and robust simulations that capture all nuclear degrees of freedom and weak nonadiabatic effects. The models are publicly available and constitute the current state-of-the-art in simulating reactive hydrogen chemistry at copper.  The two EFT models represent an isotropic model based on the local density friction approximation and an anisotropic model based on first-order time-dependent perturbation theory or orbital-dependent friction.

Our simulations reproduce experimentally measured quantities and show an improved description over previous theoretical studies. By resolving the dynamics across initial rovibrational states and multiple Cu surface facets, we demonstrate that dissociative chemisorption remains largely adiabatic under thermal conditions, with electronic friction contributing only modest corrections. While subtle differences between LDFA and ODF emerge in specific regimes, particularly at low translational energies and on more corrugated surfaces, mode-specific friction effects remain small in this context. Notably, vibrational promotion varies significantly between facets and plays a key role in modulating reactivity at sub-barrier collision energies.

The machine learning models presented here provide a versatile framework for probing the role of nonadiabatic effects in surface reactions. While electronically nonadiabatic dissociation of H\textsubscript{2} appears to be weak, other observables such as kinetic energy loss distributions, ultrafast light-driven desorption probabilities,~\cite{Juaristi2017, fuchsel_dissipative_2011} or kinetic isotope effects~\cite{zhang_ring_2025, litman_dissipative_2022, nitz_thermal_2024} may offer more sensitive fingerprints of nonadiabaticity. The presented ML models will support future efforts to explore such fingerprints. Our machine-learning representation of the electronic friction tensor is readily generalizable to other adsorbates, coverages, and materials, offering a path toward unified, high-resolution models of reactive gas-surface dynamics in the presence of weak nonadiabatic effects.

\section*{Data and Code Availability}
The databases, models, and MD scripts can be accessed through the GitHub repository: \url{https://github.com/wgst/ml-gas-surface.git}. 
MD simulations were performed within the publicly available open-source code NQCDynamics.jl. The source code and documentation are available on Github: \url{https://nqcd.github.io/NQCDynamics.jl}.


\section*{Acknowledgments}
    This work is partly based on W.G.S PhD thesis.~\cite{stark_machine_2024} It was financially supported by The Leverhulme Trust (RPG-2019-078), the UKRI Future Leaders Fellowship programme (MR/S016023/1 and MR/X023109/1), and a UKRI frontier research grant (EP/X014088/1). High-performance computing resources were provided via the Scientific Computing Research Technology Platform of the University of Warwick, the EPSRC-funded Materials Chemistry Consortium (EP/R029431/1, EP/X035859/1), and the UK Car-Parrinello consortium (EP/X035891/1) for the ARCHER2 UK National Supercomputing Service, and the EPSRC-funded HPC Midlands+ computing centre for access to Sulis (EP/P020232/1). We acknowledge helpful discussions with Alexander Spears.


\providecommand{\latin}[1]{#1}
\makeatletter
\providecommand{\doi}
  {\begingroup\let\do\@makeother\dospecials
  \catcode`\{=1 \catcode`\}=2 \doi@aux}
\providecommand{\doi@aux}[1]{\endgroup\texttt{#1}}
\makeatother
\providecommand*\mcitethebibliography{\thebibliography}
\csname @ifundefined\endcsname{endmcitethebibliography}
  {\let\endmcitethebibliography\endthebibliography}{}

\end{document}


\title{Supplemental Material for \\``Nonadiabatic reactive scattering of hydrogen on different surface facets of copper''}

\author{Wojciech G. Stark}
\address{Department of Chemistry, University of Warwick, Gibbet Hill Road, Coventry CV4 7AL, United Kingdom}
\author{Connor L. Box}
\address{Department of Chemistry, University of Warwick, Gibbet Hill Road, Coventry CV4 7AL, United Kingdom}
\author{Matthias Sachs}%
\affiliation{School of Mathematics, University of Birmingham, Birmingham, United Kingdom}
\author{Nils Hertl}
\address{Department of Chemistry, University of Warwick, Gibbet Hill Road, Coventry CV4 7AL, United Kingdom}
\address{Department of Physics, University of Warwick, Gibbet Hill Road, Coventry CV4 7AL, United Kingdom}
\author{Reinhard J. Maurer}%
\email{r.maurer@warwick.ac.uk}
\address{Department of Chemistry, University of Warwick, Gibbet Hill Road, Coventry CV4 7AL, United Kingdom}
\address{Department of Physics, University of Warwick, Gibbet Hill Road, Coventry CV4 7AL, United Kingdom}

\date{\today} 
\maketitle

\tableofcontents

\clearpage

\section{Calculation of dynamical observables}
    The sticking probability ($P_{\mathrm{sticking}}$) describes the probability that an incident gas molecule dissociates onto a surface upon collision. In this study, the sticking probability is calculated from
    \begin{equation} \label{eq:t_sticking}
         P_{\mathrm{sticking}} = \frac{n_{\mathrm{dissoc}}}{n_{\mathrm{trajs}}} ,
    \end{equation}
    where $n_{\mathrm{dissoc}}$ refers to the number of MD trajectories that ended with the dissociation of a molecule, and $n_{\mathrm{trajs}}$ to the number of all MD trajectories, including the trajectories that ended with both dissociation and scattering of a molecule.
    
    Both survival and transition probabilities relate to the rovibrational states of the gas molecule before and after scattering at the Cu surface. Survival probability refers to the probability of elastic scattering, meaning that a rovibrational state of the molecule did not change after the collision with the surface, and can be evaluated according to
    \begin{equation} \label{eq:t_sticking}
         P_{\mathrm{survival}} = \frac{n_{\mathrm{surv}}}{n_{\mathrm{trajs}}} ,
    \end{equation}
    where $n_{\mathrm{surv}}$ refers to the number of rovibrationally elastic MD trajectories.
    
    Transition probability refers to the probability of inelastic scattering, which includes a certain (de-)excitation of the vibrational and/or rotational state of the molecule.
    \begin{equation} \label{eq:t_sticking}
         P_{\mathrm{transition}} = \frac{n_{\mathrm{trans}}^{\mathrm{\nu}_{0},\mathrm{J}_{0} \rightarrow \mathrm{\nu}_{f},\mathrm{J}_{f}}}{n_{\mathrm{trajs}}} ,
    \end{equation}
    where $n_{\mathrm{trans}}^{\mathrm{\nu}_{0},\mathrm{J}_{0} \rightarrow \mathrm{\nu}_{f},\mathrm{J}_{f}}$ is the number of MD trajectories in which molecule of $\mathrm{\nu}_{0},\mathrm{J}_{0}$ initial rovibrational state inelastically scattered from surface with the final rovibrational state of $\mathrm{\nu}_{f},\mathrm{J}_{f}$.

    \subsection{Statistical error evaluation}
        To ensure that the evaluated dynamical observables are converged, we have run 20,000 trajectories for every data point shown in the main manuscript (every surface and collision energy). To confirm that the results are well converged, we evaluated statistical error using a bootstrapping method, in a similar fashion as for evaluation of sticking probability in our previous study~\cite{stark_machine_2023}. In this approach, we separate 20,000 results into groups containing 500, 1,000, 2,000, 2,500, 4,000, or 5,000 data points, in such a fashion that the first group contains 40 subgroups of 500 data points, the second group 20 subgroups of 1,000 data points, and so on. Such separation is done 50 times for every group, and inverse variance is obtained for all the groups. Using linear regression, the error for the final group size (20,000) can be read. Resulting error bars for sticking and survival probabilities are negligible. The highest errors can be observed for transition probabilities, however, even in this case, the errors are negligible. In Fig.~\ref{fig:results_sts_v2J1_err}, transition probabilities are shown for state-to-state scattering of H\textsubscript{2}($\mathrm{\nu}$=2,~J=1~$\rightarrow$~$\mathrm{\nu}$=1,~J=1) at different Cu surfaces using BOMD, MDEF-LDFA, and MDEF-ODF methods (just as in Fig. 7), including error bars. In all the cases, evaluated statistical errors do not exceed 0.06\%.


\section{Coordinate Transformation of Electronic Friction Tensor} \label{sec:methods_mdef_icoors}
    Often, internal coordinates are used to represent the electronic friction in diatomic adsorbates, which enables a better understanding of the directional dependence of EFT. Both Cartesian (x, y, z) and internal (d, $\theta$, $\phi$, X, Y, Z) coordinates are shown schematically in Fig.~1. In the internal coordinate system, d is the distance between adsorbate atoms of a diatomic molecule, $\theta$ and $\phi$ are the polar and azimuthal angles. Finally, the X, Y, and Z are the coordinates of the center of mass of the adsorbate atoms. The transformation of the EFT, $\bm{\Lambda}$, from the Cartesian to internal coordinates is as follows
    \begin{equation} \label{eq:int_coords_transf}
        \bm{\Lambda}_{\mathrm{int}} = \bm{U}^T \bm{\Lambda}_{\mathrm{cart}} \bm{U},
    \end{equation}
    where $\bm{U}$ is the transformation matrix between the coordinate systems of the diatomic molecule
    \begin{equation} \label{eq:int_coords}
        \bm{U} =
            \begin{bmatrix}
                \frac{x_1 - x_2}{r} & \frac{(x_2 - x_1)(z_2 - z_1)}{r^2 r'} & \frac{y_1 - y_2}{r'^2} & m_1/M & 0 & 0\\
                \frac{y_1 - y_2}{r} & \frac{(y_2 - y_1)(z_2 - z_1)}{r^2 r'} & \frac{x_2 - x_1}{r'^2} & 0 & m_1/M & 0\\
                \frac{z_1 - z_2}{r} & \frac{-r'}{r^2} & 0 & 0 & 0 & m_1/M\\
                \frac{x_2 - x_1}{r} & \frac{(x_1 - x_2)(z_2 - z_1)}{r^2 r'} & \frac{y_2 - y_1}{r'^2} & m_2/M & 0 & 0\\
                \frac{y_2 - y_1}{r} & \frac{(y_1 - y_2)(z_2 - z_1)}{r^2 r'} & \frac{x_1 - x_2}{r'^2} & 0 & m_2/M & 0\\
                \frac{z_2 - z_1}{r} & \frac{r'}{r^2} & 0 & 0 & 0 & m_2/M\\
            \end{bmatrix},
    \end{equation}
    where $x_n$, $y_n$, $z_n$ are Cartesian coordinates, and $m_n$ is the mass of the $n$-th hydrogen atom,\\
    $r = \sqrt{(x_2 - x_1)^2 + (y_2 - y_1)^2 + (z_2 - z_1)^2}$ and $r' = \sqrt{(x_2 - x_1)^2 + (y_2 - y_1)^2}$, and $M = m_1 + m_2$. The column vectors of the transformation matrix are normalized to conserve the trace of the friction tensor during projection. Note that such normalization was not done in Fig.~S13 to enable direct comparison with the referenced works, which did not include such a procedure in their coordinate transformation scheme. The resulting EFT contains the diagonal values, which are associated with the electronic friction in the internal coordinates (d, $\theta$, $\phi$, X, Y, Z), and non-diagonal values that correspond to the electronic friction coupled between two of the internal coordinates.


\newpage
\section{Local Density Friction Approximation (ACE) Model} \label{sec:eft_models_ldfa}
\begin{figure}[h!]
        \centering
        \includegraphics[width=0.55\linewidth]{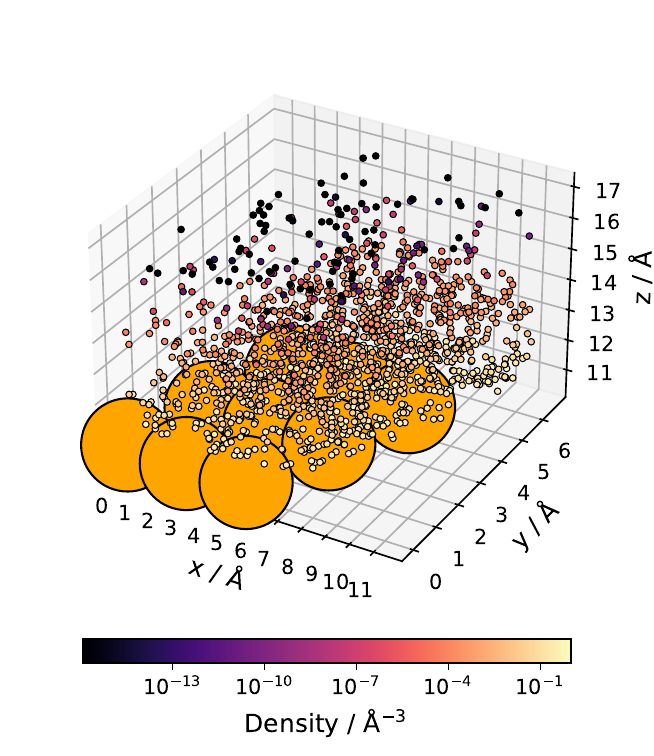}
        \caption{\textbf{Distribution of virtual hydrogen atoms at Cu(111) surface in the final LDFA (density) data set.} The plot contains all hydrogen atoms in the data set used for the construction of the density model (every structure contains 1 H atom and 54 Cu atoms). The top layer of the copper surface is shown schematically in a relaxed position.}
        \label{fig:ldfa_train_data}
    \end{figure}

\begin{figure*}[h!]
            \centering
            \includegraphics[width=0.7\linewidth]{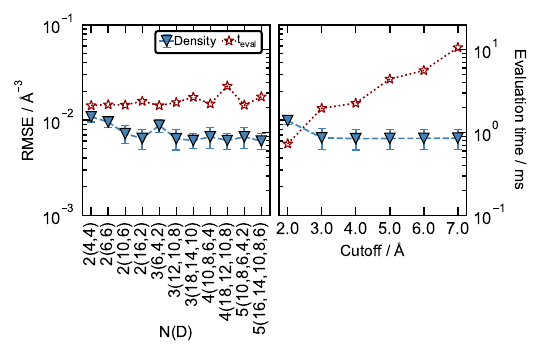}
            \caption[Optimization of ACE-based density model parameters.]{\textbf{Optimization of ACE-based density model parameters.} Convergence of the model with respect to maximum correlation order (N), and corresponding maximum polynomial degree (D) (left) and cutoff distance (right), based on the density (blue triangles) test RMSEs ($\mathrm{\AA}^{-3}$). The error bars represent the standard deviation between the RMSEs obtained for all the splits in cross-validation. Additionally, corresponding average model evaluation times, based on 100 model evaluations, are included (red stars). All the evaluation times have been obtained using a single CPU core, AMD EPYC 7742 (Rome).}
            \label{fig:opt_ldfa_ace}
        \end{figure*}

\begin{figure*}[h!]
            \centering
            \includegraphics[width=0.35\linewidth]{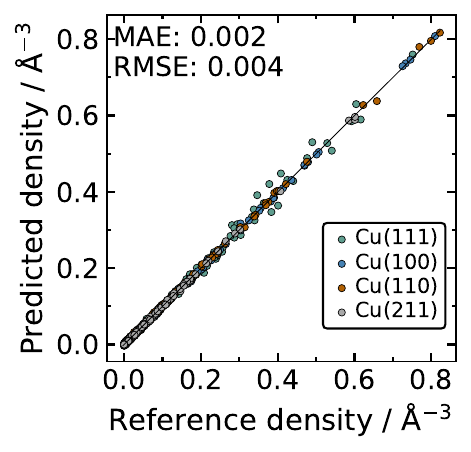}
            \caption[Performance of the ACE model in predicting electron densities of clean Cu surfaces at the positions of hydrogen atoms.]{\textbf{Performance of the ACE model in predicting electron densities of clean Cu surfaces at the positions of hydrogen atoms.} The ACE density model test set predictions for all 4 Cu facets compared to the reference values obtained with FHI-aims.}
            \label{fig:ace_dens_pred_vs_red}
        \end{figure*}

\begin{figure}[h!]
            \centering
            \includegraphics[width=0.6\linewidth]{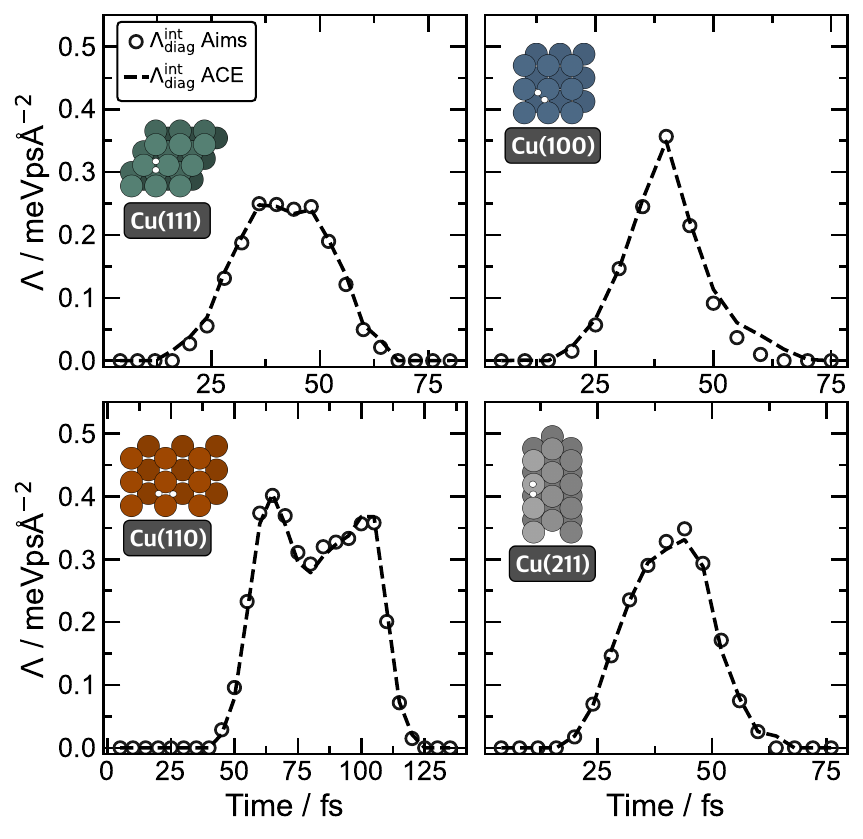}
            \caption[Predictions of LDFA-based EFT elements along the trajectories of H\textsubscript{2} scattering at Cu surfaces.]{\textbf{Predictions of LDFA-based EFT elements along the trajectories of H\textsubscript{2} scattering at Cu surfaces.} The diagonal elements of EFT (internal coordinates) are plotted for scattering trajectories at Cu surfaces, predicted by the ACE model, and compared to reference values calculated using FHI-aims. Note that all the diagonal components of EFT are identical in normalized internal coordinates.}
            \label{fig:ldfa_model_val_scatter}
        \end{figure}

    The surrogate model of electronic friction based on the Atomic Cluster Expansion (ACE) in the local density friction approximation (LDFA) was trained on bare surface density at the position of a hydrogen atom. The training data includes many density readouts at different positions above the four Cu surface facets as shown in Fig.~\ref{fig:ldfa_train_data}. Many such readouts were made for 200 different surface configurations of 4 different Cu facets at one of 3 temperatures 300, 600, or 900~K.
        Details of the hyperparameter optimization of the ACE-based electron density model are shown in Fig.~\ref{fig:opt_ldfa_ace}.
        The density model performance is shown in Fig.~\ref{fig:ace_dens_pred_vs_red}.
        In Fig.~\ref{fig:ldfa_model_val_scatter}, the performance of the LDFA EFT model is shown along scattering (Fig.~\ref{fig:ldfa_model_val_scatter}) trajectories for all four considered Cu facets, compared with DFT-based calculations.
        
\clearpage

\section{Orbital-Dependent Friction Convergence} \label{sec:eft_odf_convergence}

\begin{figure}[h!]
        \centering
        \includegraphics[width=0.6\linewidth]{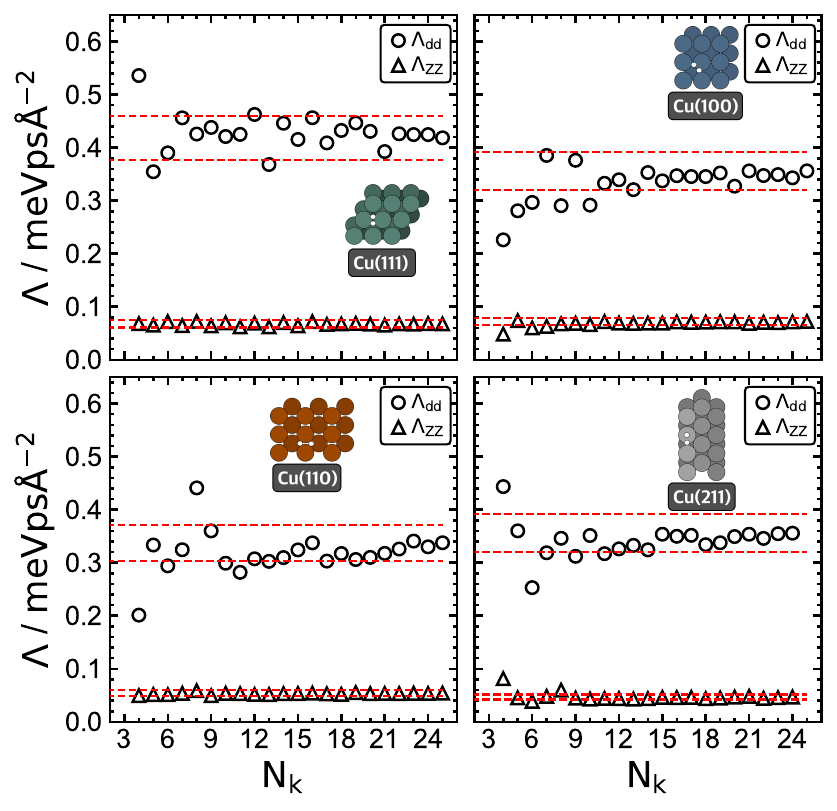}
        \caption{\textbf{Optimization of k-grid size with respect to EFT elements.} Convergence of two EFT elements (dd and ZZ) with respect to k-grid of size N\textsubscript{k}$\times$N\textsubscript{k}$\times$1 for the transition state (TS) structure of H\textsubscript{2} dissociation at Cu surfaces. 3$\times$3 slabs (1$\times$4 for Cu(211)) with 6 layers of Cu and a Gaussian-type broadening $\sigma=0.6$ were used in calculations. Red, dashed lines represent $\pm10\%$ deviation of the value at the highest N\textsubscript{k}.}
        \label{fig:conv_eft_kgrid}
    \end{figure}
\begin{figure}[h!]
        \centering
        \includegraphics[width=0.6\linewidth]{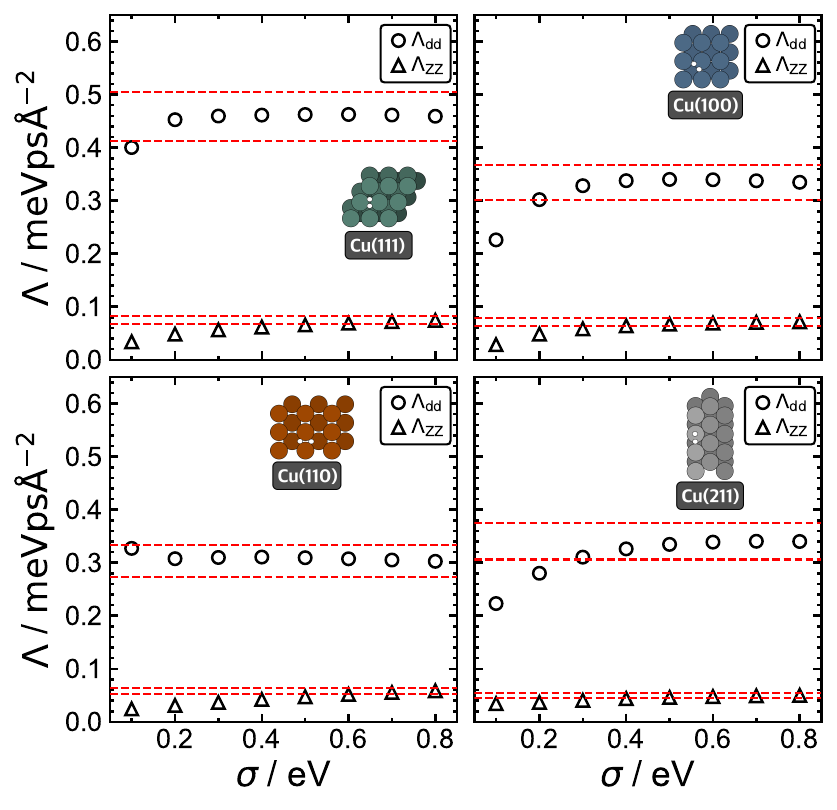}  
        \caption[Optimization of broadening parameter with respect to EFT elements.]{\textbf{Optimization of broadening parameter with respect to EFT elements.} Convergence of two EFT elements (dd and ZZ) with respect to Gaussian-type broadening ($\sigma$) for the TS structure of H\textsubscript{2} dissociation at Cu surfaces. 3$\times$3 slabs (1$\times$4 for Cu(211)) with 6 layers of Cu and a k-grid of size 12$\times$12$\times$1 were used in calculations. Red, dashed lines represent $\pm10\%$ deviation of the value at the highest $\sigma$.}
        \label{fig:conv_eft_broadening}
    \end{figure}
\begin{figure}[h!]
        \centering
        \includegraphics[width=0.6\linewidth]{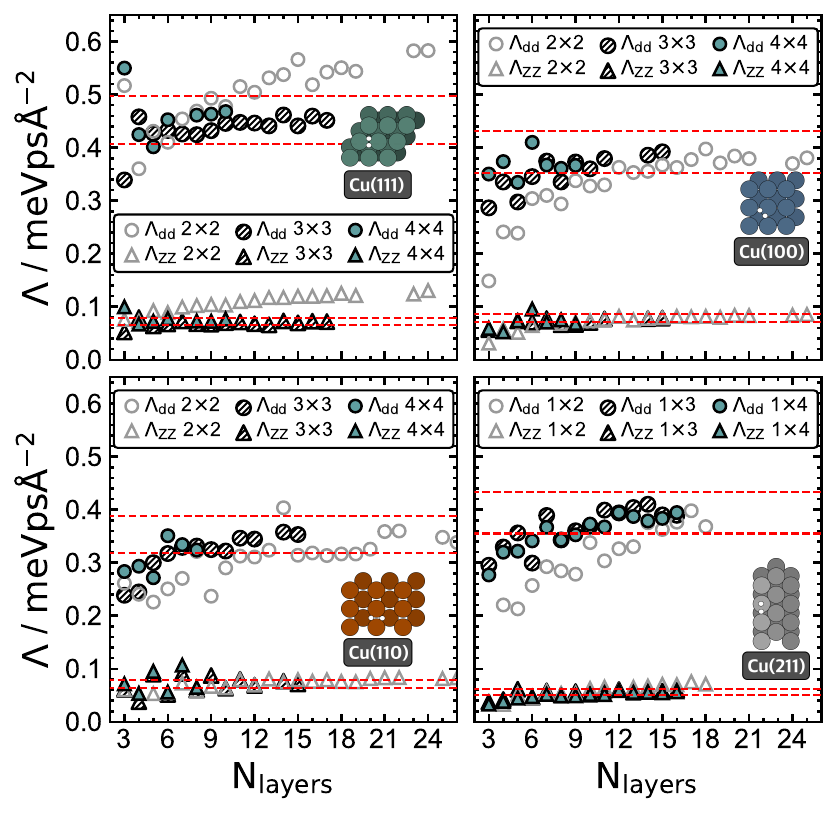}
        \caption{\textbf{Convergence of slab size with respect to EFT elements.} Convergence of two EFT elements (dd and ZZ) with respect to the number of layers (N\textsubscript{layers}) and unit cell size at the TS of H\textsubscript{2} dissociative chemisorption at Cu surfaces. A Gaussian-type broadening $\sigma=0.6$ and a k-grid of size 12$\times$12$\times$1 were used in all calculations. Red, dashed lines represent $\pm10\%$ deviation of the value corresponding to the highest N\textsubscript{layers} of the 3$\times$3 (4$\times$1 in the case of Cu(211)) unit cell.}
        \label{fig:conv_eft_layers}
    \end{figure}

    The orbital-dependent friction EFT convergence with respect to k-grid size, broadening parameter, $\sigma$, and number of layers in the slab, $N_\mathrm{layers}$, are shown in Figs.~\ref{fig:conv_eft_kgrid}, ~\ref{fig:conv_eft_broadening},~and~\ref{fig:conv_eft_layers}, respectively.
    
    
\clearpage

\section{Orbital-Dependent Friction (ACE-friction) Model} \label{sec:eft_models_odf}

    \subsection{Parameter Optimization}
\begin{figure*}
            \centering
            \includegraphics[width=0.7\linewidth]{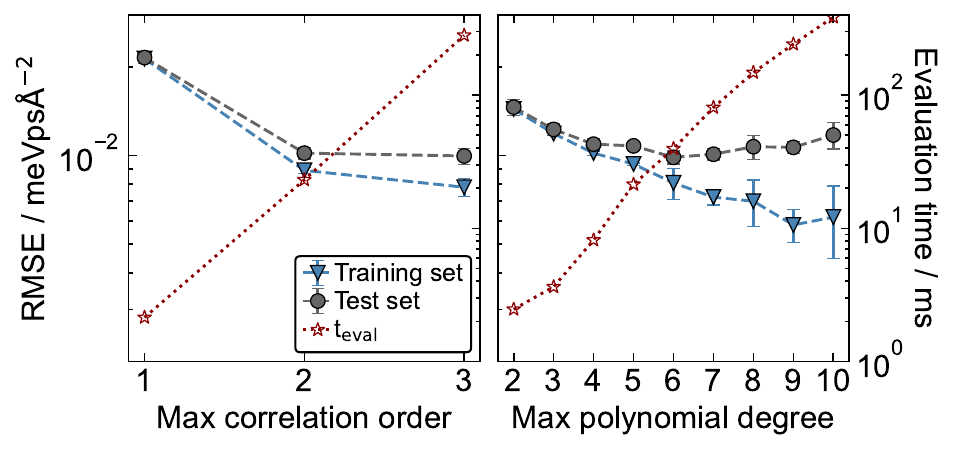}
            \caption{\textbf{Optimization of ACE-friction model parameters: correlation order and polynomial degree.} Convergence of RMSE with respect to maximum correlation order (left) and maximum polynomial degree (right), based on the training (blue triangles) and test (grey circles) RMSEs of electronic friction (meVps$\mathrm{\AA}^{-2}$). The error bars represent the standard deviation between the RMSEs obtained for all the splits in cross-validation. Additionally, corresponding average model evaluation times, based on 100 model evaluations, are included (red stars). The evaluation times were obtained using a single CPU Intel Xeon Platinum 826 (Cascade Lake) 2.9 GHz processor core.}
            \label{fig:opt_acefriction_correlation}
        \end{figure*}

\begin{figure*}[h!]
            \centering
            \includegraphics[width=0.5\linewidth]{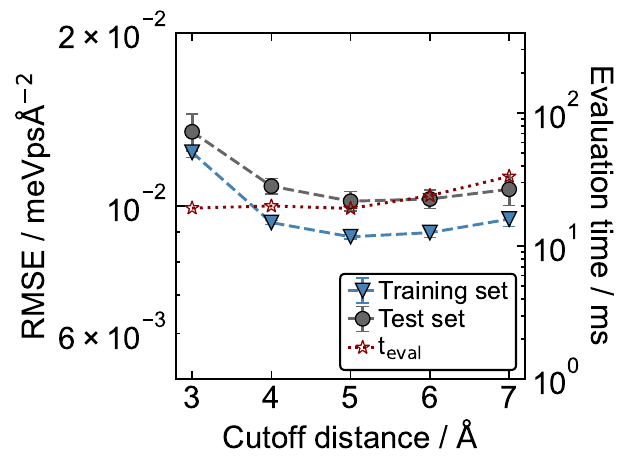}
            \caption{\textbf{Optimization of ACE-friction model parameters: cutoff distance.} Convergence of the model with respect to cutoff distance ($\mathrm{\AA}$), based on the training (blue triangles) and test (grey circles) RMSEs of electronic friction (meVps$\mathrm{\AA}^{-2}$). The error bars represent the standard deviation between the RMSEs obtained for all the splits in cross-validation. Additionally, corresponding average model evaluation times, based on 100 model evaluations, are included (red stars). The evaluation times were obtained using a single CPU Intel Xeon Platinum 826 (Cascade Lake) 2.9 GHz processor core.}
            \label{fig:opt_acefriction_cutoff}
        \end{figure*}

        Optimization of ACE-friction model parameters is shown in Figs.~\ref{fig:opt_acefriction_correlation}~and~\ref{fig:opt_acefriction_cutoff}.
        
        
    \subsection{Model Validation}
\begin{figure*}[h!]
            \centering
            \includegraphics[width=0.4\linewidth]{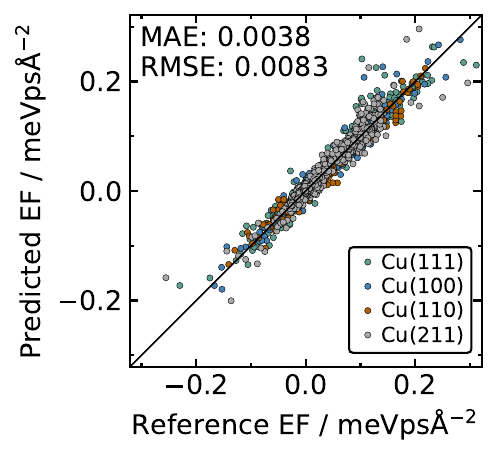}
            \caption[Performance of the ACE-friction model in predicting all electronic friction Cartesian elements for H\textsubscript{2}/Cu structures.]{\textbf{Performance of the ACE-friction model in predicting all electronic friction Cartesian elements for H\textsubscript{2}/Cu structures.} The ACE-friction model predictions compared to the reference EFT elements (in Cartesian coordinates) for the test set. Each EFT contains 6$\times$6 (36) values, including 6 diagonal elements. The negative values correspond only to non-diagonal EFT elements.}
            \label{fig:aceds_pred_vs_red}
        \end{figure*}

        The performance of the final ACE-friction model is shown in Fig.~\ref{fig:aceds_pred_vs_red}.
    
\begin{figure}[h!]
            \centering
            \includegraphics[width=0.6\linewidth]{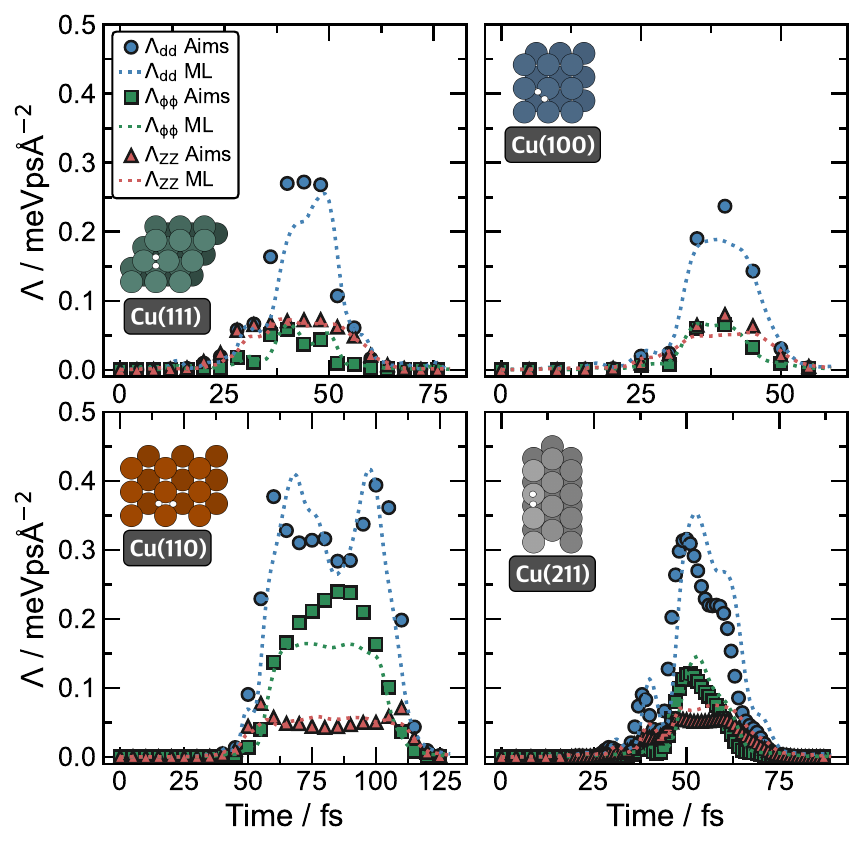}
            \caption[Predictions of ODF-based EFT elements across trajectory steps of H\textsubscript{2} scattering at Cu surfaces.]{\textbf{Predictions of ODF-based EFT elements across trajectory steps of H\textsubscript{2} scattering at Cu surfaces.} Electronic friction values (internal coordinates) are plotted for scattering trajectories at Cu surfaces predicted by the ACE-friction model and compared to reference values calculated using FHI-aims.}
            \label{fig:odf_model_val_scatter}
        \end{figure}

        The ODF-based EFT was evaluated with the ACE-friction model along a scattering trajectory, and compared with DFT-based reference calculations (Fig.~\ref{fig:odf_model_val_scatter}).

    \subsection{Learning Behaviour}
        \begin{figure*}[h!]
            \centering
            \includegraphics[width=0.4\linewidth]{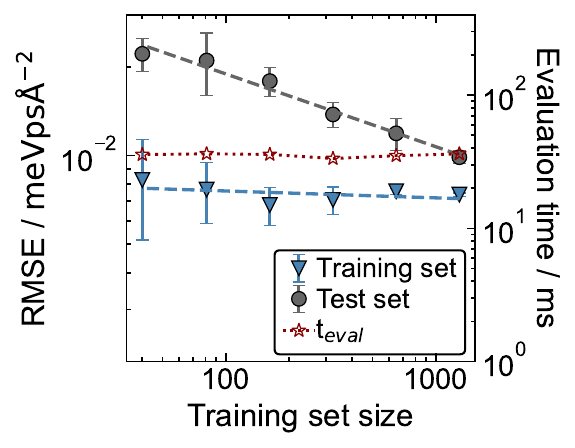}
            \caption[Learning curve for ACE-friction EFT model.]{\textbf{Learning curve for ACE-friction EFT model.} Log-log plot of the averaged test set predictive error (RMSE) of EFT elements as a function of training set size for ACE-friction model. The shown RMSE values are averaged over 5 cross-validation splits, excluding validation sets, differing only by training sets. Error bars correspond to standard deviations between the RMSEs obtained over all splits. Red stars indicate an average evaluation time of an entire EFT based on 100 evaluations.}
            \label{fig:acefriction_learning_curve}
        \end{figure*}
        
        The learning curve, showing the electronic friction test set RMSEs for models trained on data sets with different sizes, is shown in Fig.~\ref{fig:acefriction_learning_curve}. As expected, the test set RMSEs are improving with more training data, reaching the proximity of the training set RMSEs with roughly 1300 structures. There is no visible improvement in the prediction of training set RMSEs across the models with different training set sizes. This could be associated with the very early convergence of the training set error and further improvement being stopped by the numerical noise and intrinsic error of the electronic friction evaluation with the electronic structure code. The relatively high intrinsic error, thus the lack of smoothness of electronic friction elements, was reported before by Zhang~\textit{et~al.} for H\textsubscript{2} on Ag(111)~\cite{zhang_symmetry-adapted_2020}.
        
\clearpage

\section{Electronic friction tensor elements along minimum energy paths compared to previous theoretical results} \label{sec:h2cu_neb_ref}

    \begin{figure}[h!]
        \centering
        \includegraphics[width=0.9\linewidth]{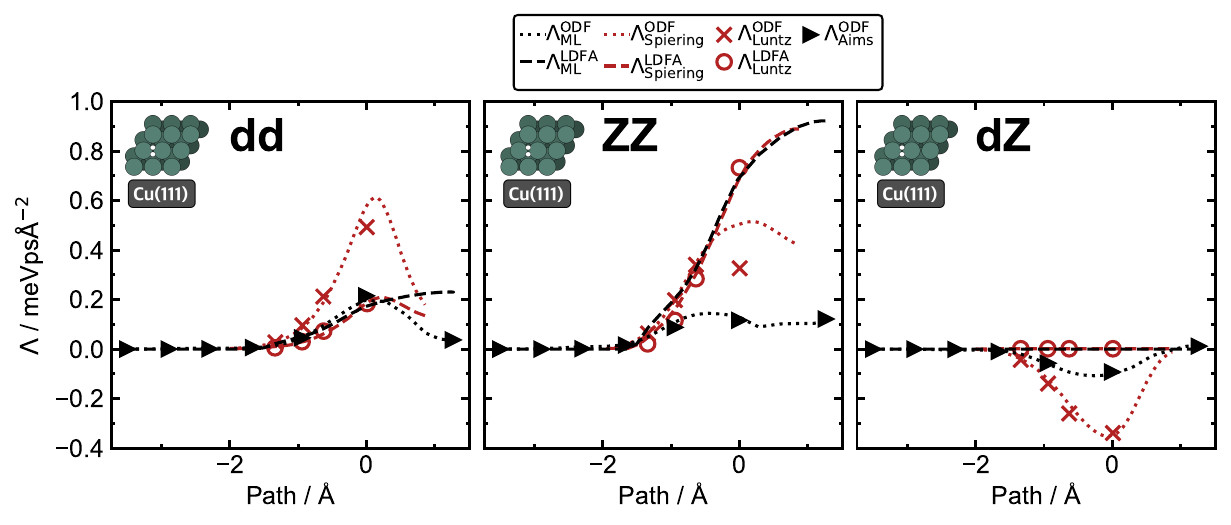}
        \caption{\textbf{Electronic friction tensor elements along minimum energy path of H\textsubscript{2} dissociation at Cu(111).} EFT elements, dd (left), ZZ (middle), and dZ (right) were evaluated using TDPT code within FHI-aims, and ML models constructed in this study (LDFA and ODF) and compared with the previous theoretical results by Spiering~and~Meyer~\cite{spiering_testing_2018} and Luntz~\textit{et~al.}~\cite{luntz_comment_2009,luntz_how_2005}. Pictures in the upper-left corners depict the transition state structures of the minimum energy path.}
        \label{fig:h2cu_neb_cu111_ref}
    \end{figure}

    EFT values along the minimum energy path of H\textsubscript{2}/Cu(111) dissociation calculated in this study are compared with the previous theoretical results by Spiering~and~Meyer~\cite{spiering_testing_2018} and Luntz~\textit{et~al.}~\cite{luntz_comment_2009,luntz_how_2005}. The previous theoretical results did not apply normalization after the Cartesian to internal coordinate transformation, thus, here we compare non-normalized EFT values in the internal coordinate system.

    The agreement between EFT-LDFA values between all methods is close, however, significant differences can be noticed for EFT-ODF values, which is a result of the use of a different EFT equation (reference studies employ low-temperature ODF equation - Eq. 3 in the main manuscript - whereas we utilize the newer equation by Maurer~\textit{et~al.}~\cite{maurer_ab_2016} - Eq. 2 in the main manuscript). The discrepancies between ODF values obtained from the two different ODF formulations were explained in more detail by Box~\textit{et~al.}~\cite{box_ab_2023}.


\clearpage

\section{2D PES elbow plot} \label{sec:2dpes}

\begin{figure}[h!]
        \centering
        \includegraphics[width=0.6\linewidth]{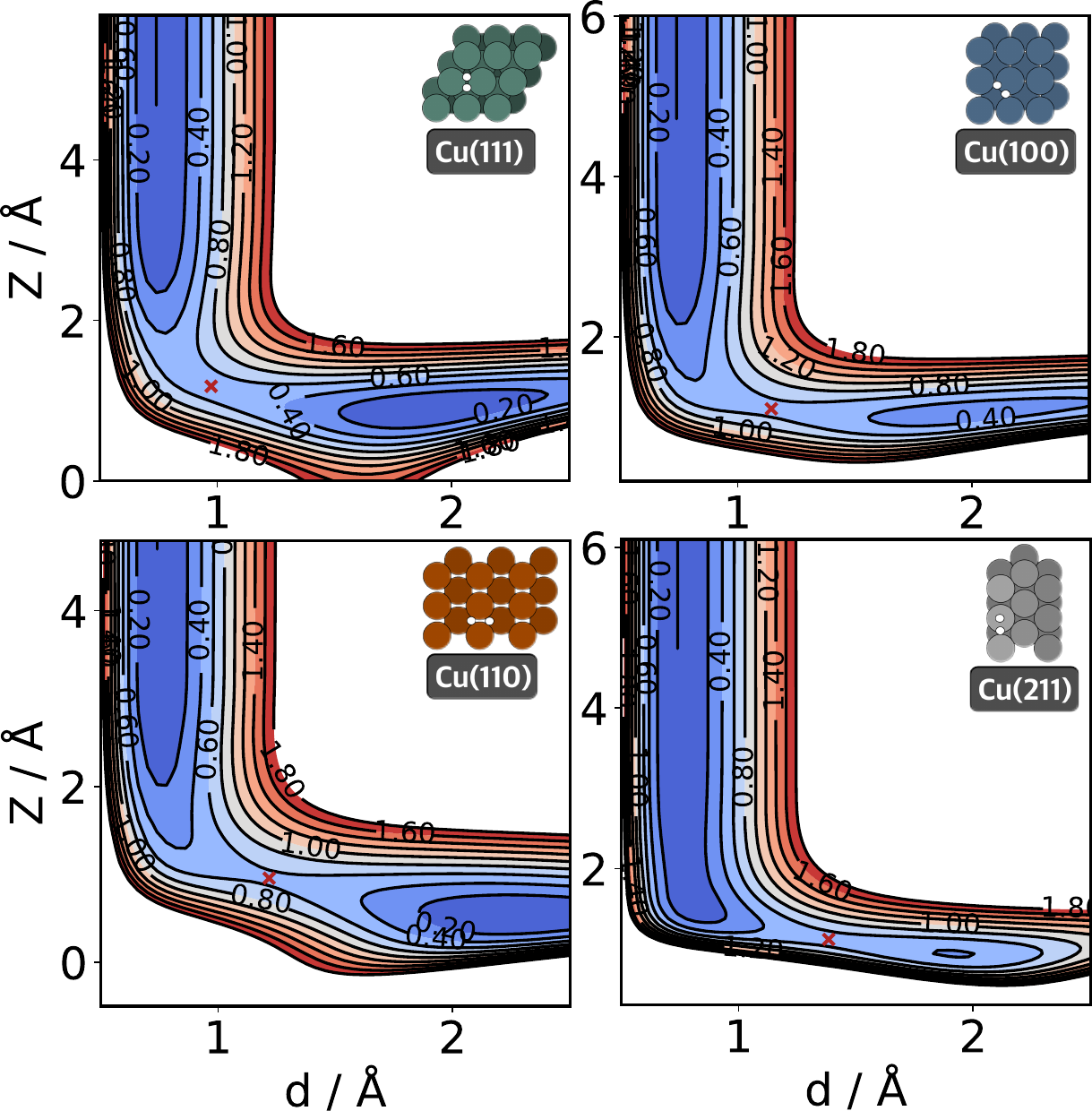}
        \caption{\textbf{Minimum energy paths obtained using CI-NEB method for H\textsubscript{2} dissociative adsorption on different copper surfaces.} Potential energy values are shown along the reaction path (\AA), evaluated using MACE PES from Ref.~\cite{stark_benchmarking_2024}. The transition state is marked with a red cross.}
        \label{fig:si_elbows}
    \end{figure}

    2D projection of PES is shown in Fig.~\ref{fig:si_elbows} along the reaction path (distance between hydrogen atoms and distance between hydrogen molecule and the top layer of metal surface) for H\textsubscript{2} dissociation at different Cu surfaces.
    
\clearpage

\section{Fingerprint for nonadiabatic effects in state-to-state scattering}
        \begin{figure}
            \centering
            \includegraphics[width=0.75\linewidth]{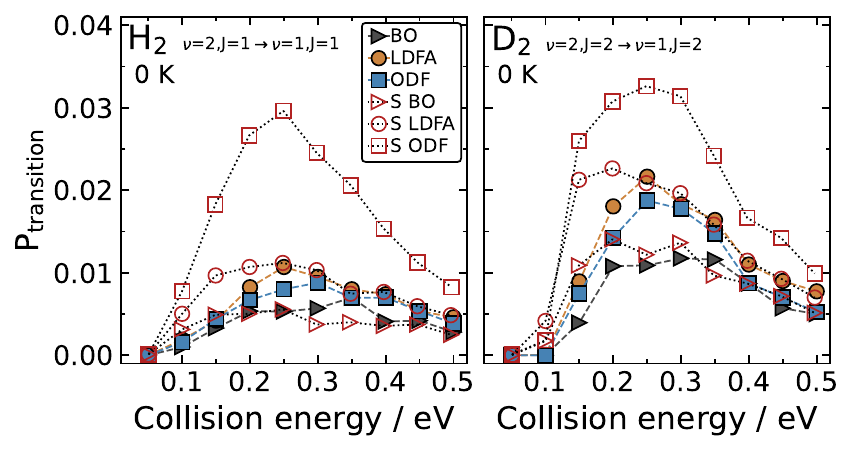}
            \caption{\textbf{Vibrational de-excitation probabilities for state-to-state scattering of H\textsubscript{2} and D\textsubscript{2} at Cu(111) (0~K).} Vibrational de-excitation of H\textsubscript{2}($\mathrm{\nu}$=2,~J=1~$\rightarrow$~$\mathrm{\nu}$=1,~J=1, left) and D\textsubscript{2}($\mathrm{\nu}$=2,~J=2~$\rightarrow$~$\mathrm{\nu}$=1,~J=2, right) at Cu(111) surface are plotted for a set of translational energies (eV). The probabilities were calculated using classical BOMD (triangles), MDEF-LDFA (circles), and MDEF-ODF (squares) and compared with the theoretical results obtained by Spiering~and~Meyer~\cite{spiering_testing_2018}, indicated by the letter ``S'' (empty markers).}
            \label{fig:results_sts_h2d2cu111}
        \end{figure}

        The rovibrational state-to-state scattering of H\textsubscript{2} from Cu(111) has previously been identified as a sensitive fingerprint for nonadiabatic effects~\cite{spiering_testing_2018},  primarily because the rotational quantum number constrains the scattering into a narrow final-state distribution.
        Thus we revisit vibrational de-excitation probabilities for H\textsubscript{2}($\mathrm{\nu}$=2,~J=1~$\rightarrow$~$\mathrm{\nu}$=1,~J=1) and D\textsubscript{2}($\mathrm{\nu}$=2,~J=2~$\rightarrow$~$\mathrm{\nu}$=1,~J=2) on Cu(111). 

        All results (Fig.~\ref{fig:results_sts_h2d2cu111}) show a general behavior of de-excitation probabilities increasing with collision energy, peaking near 0.25~eV, then declining. Spiering and Meyer identified this characteristic shape as the interplay between increasing approach to the surface, increasing the vibrational de-excitation probability, with the dissociation channel becoming more effective at higher collision energies. 
        Our LDFA results align with those of Spiering and Meyer, except for a discrepancy at 0.1--0.2 eV, also observed in BOMD simulations. This discrepancy likely arises from differences in the employed PESs, as Spiering and Meyer employed the SRP48 functional and the static surface approximation.

        Spiering and Meyer found that de-excitation probabilities at 0.25~eV using ODF ($\Lambda^{0}$) were three times larger than when using LDFA, highlighting directional dependence in electronic friction. However, when the low-temperature approximation is not applied ($\Lambda$, Eq.~\ref{eq:odf_sd}), we observe close agreement between MDEF-ODF and MDEF-LDFA across all energies (Fig.~\ref{fig:results_sts_h2d2cu111}). This aligns with Maurer~\textit{et~al.}~\cite{maurer_hot_2019}, who reported no significant differences between ODF ($\Lambda$) and LDFA in the de-excitation probabilities of H\textsubscript{2}($\mathrm{\nu}$=2,~J=0~$\rightarrow$~$\mathrm{\nu}$=1,~J=0) on Ag(111).

        We find a more profound enhancement of the peak upon applying electronic friction for D$_2$ compared to H\textsubscript{2}, the difference between employing LDFA or ODF remains minimal, however. 
        Our results suggest that the anisotropy is not significant for vibrational de-excitation probabilities for H\textsubscript{2} or D\textsubscript{2} scattering on Cu(111) at 0 K, and we do not expect these differences to be experimentally resolvable.

\clearpage
\section{Intersection between electronic friction and velocity in scattering} \label{sec:eft_vs_vel}
\begin{figure}[h!]
        \centering
        \includegraphics[width=0.7\linewidth]{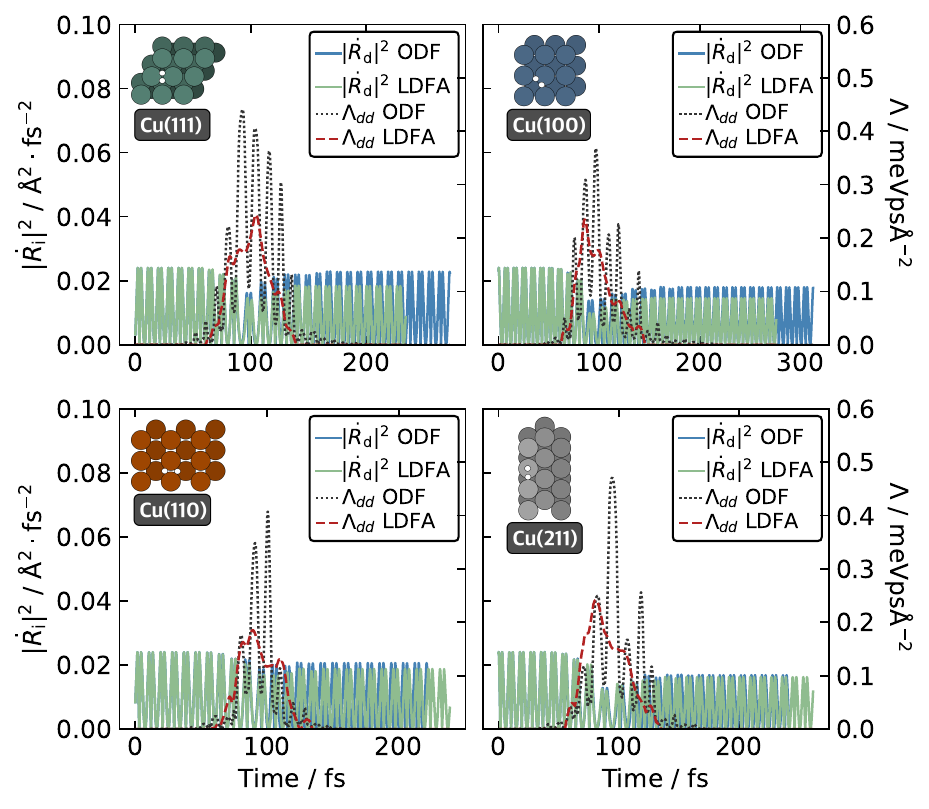}
        \caption{\textbf{Square of velocity in d direction along the H\textsubscript{2} scattering trajectories at different Cu surfaces together with corresponding electronic friction values.} Square of velocity projected along d direction ($|\dot{R}_{\mathrm{d}}|^{2}$ and dd elements of EFT ($\Lambda_{\mathrm{dd}}$) were evaluated for a scattering trajectory ($\mathrm{\nu}$=2,~J=1) calculated with LDFA and ODF for a collision energy of 0.45~eV.}
        \label{fig:results_sts_vel_sq_d}
    \end{figure}

\begin{figure}[h!]
        \centering
        \includegraphics[width=0.7\linewidth]{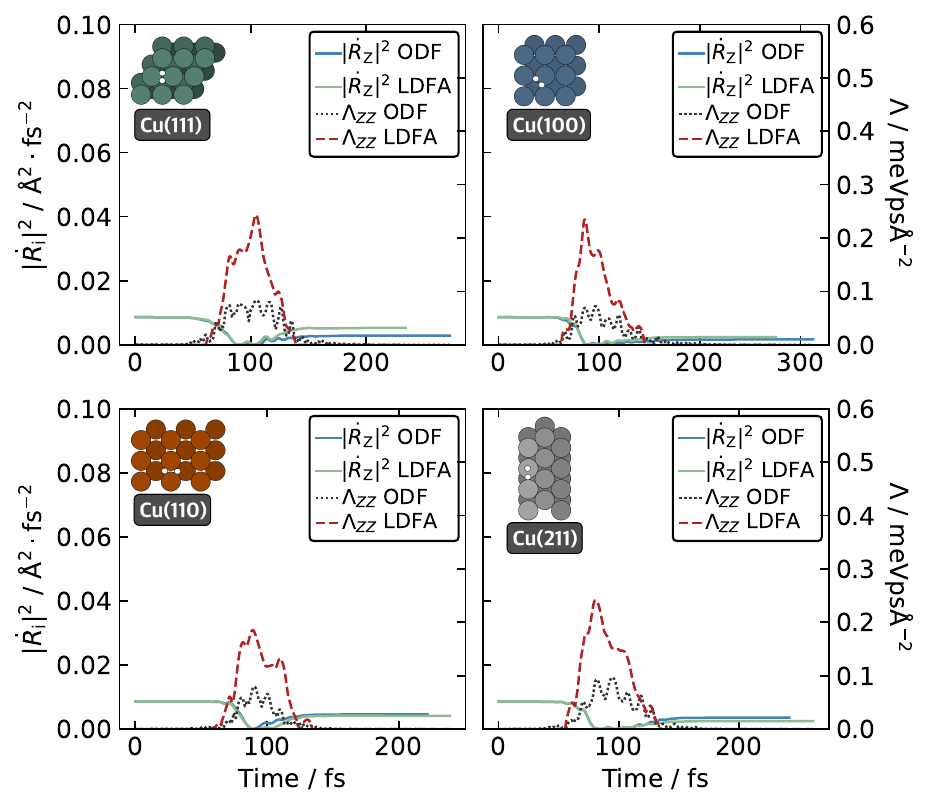}
        \caption[Square of velocity in Z direction along the H\textsubscript{2} scattering trajectories at different Cu surfaces together with corresponding electronic friction values.]{\textbf{Square of velocity in Z direction along the H\textsubscript{2} scattering trajectories at different Cu surfaces together with corresponding electronic friction values.} Square of velocity projected along Z direction ($|\dot{R}_{\mathrm{Z}}|^{2}$ and ZZ elements of EFT ($\Lambda_{\mathrm{ZZ}}$) were evaluated for a scattering trajectory ($\mathrm{\nu}$=2,~J=1) calculated with LDFA and ODF for a collision energy of 0.45~eV.}
        \label{fig:results_sts_vel_sq_z}
    \end{figure}

    The square of velocity and EFT along MD trajectory in d and Z directions were plotted in Figs~\ref{fig:results_sts_vel_sq_d}~and~\ref{fig:results_sts_vel_sq_z}, respectively.
    
\section{Translational energy gain}

        \begin{figure}
            \centering
            \includegraphics[width=0.45\linewidth]{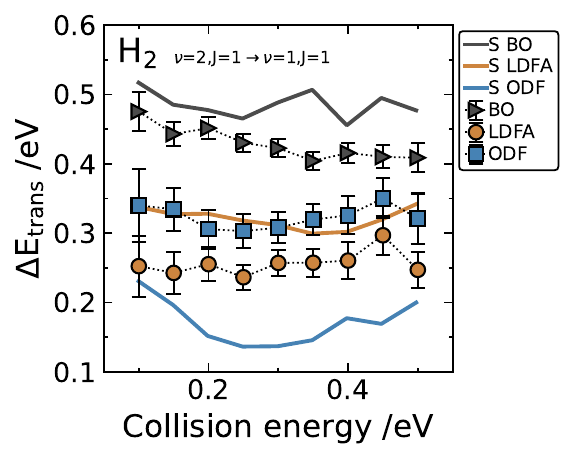}
             \caption{\textbf{Average translational energy gain at different collision energies for H\textsubscript{2}($\mathrm{\nu}$=2,~J=1~$\rightarrow$~$\mathrm{\nu}$=1,~J=1) scattering from Cu(111).} The results were obtained with adiabatic MD (BOMD) (gray), and MDEF using both LDFA (orange) and ODF (blue) EFT implementations. Solid lines represent results obtained by Spiering and Meyer~\cite{spiering_testing_2018}. More information about obtaining values for error bars is provided below.}
            \label{fig:results_eloss_cu111}
        \end{figure}

        We compare the average translational energy gain for H\textsubscript{2}($\mathrm{\nu}$=2,~J=1~$\rightarrow$~$\mathrm{\nu}$=1,~J=1) scattering from Cu(111) using BOMD, LDFA-based MDEF, and ODF-based MDEF (Fig.~\ref{fig:results_eloss_cu111}) with those reported by Spiering and Meyer~\cite{spiering_testing_2018}.
        BOMD shows a steady decrease in energy gain with increasing collision energy. Nonadiabatic effects reduce both the energy gain and its sensitivity to collision energy. 
        While Spiering predicts lower energy gain with $\Lambda^{0}$ ODF than LDFA, our $\Lambda$-variant ODF results show slightly higher energy gain than LDFA, with a difference of $\sim$0.1~eV across the tested energies. The impact of nonadiabatic effects diminishes at higher collision energies.

\subsection{Evaluation of errors} \label{sec:si_e_loss_err}
    In Fig.~\ref{fig:results_eloss_cu111} we show energy gain for H\textsubscript{2}($\mathrm{\nu}$=2,~J=1~$\rightarrow$~$\mathrm{\nu}$=2,~J=1) scattering at Cu(111) surface. The error bars are obtained in the same fashion as in Ref.~\cite{spiering_testing_2018} by first evaluating the standard error
    \begin{equation}
        \sigma^{\mathrm{SE}} = \frac{\sigma}{\sqrt{N}} = \frac{1}{N}\sqrt{\sum^{N}_{i}(\Delta E_{i}^{\mathrm{transl}} - \left< \Delta E^{\mathrm{transl}}\right> )^{2}},
    \end{equation}
    where N corresponds to the number of scattering trajectories of H\textsubscript{2}($\mathrm{\nu}$=2,~J=1~$\rightarrow$~$\mathrm{\nu}$=2,~J=1), $E^{\mathrm{transl}}_{i}$ is translational energy of H\textsubscript{2} in the final structure of trajectory $i$, and $\left< \Delta E^{\mathrm{transl}}\right>$ is an average translational energy of all trajectories that correspond to the transition of H\textsubscript{2}($\mathrm{\nu}$=2,~J=1~$\rightarrow$~$\mathrm{\nu}$=2,~J=1). 

    The final error bars are evaluated from $1.96\cdot \sigma^{\mathrm{SE}}$, which corresponds to the confidence level of 95\% assuming a normal distribution.
    

\section{Contour plots}
\label{sec:si_contours}

        \begin{figure}
            \centering
            \includegraphics[width=0.6\linewidth]{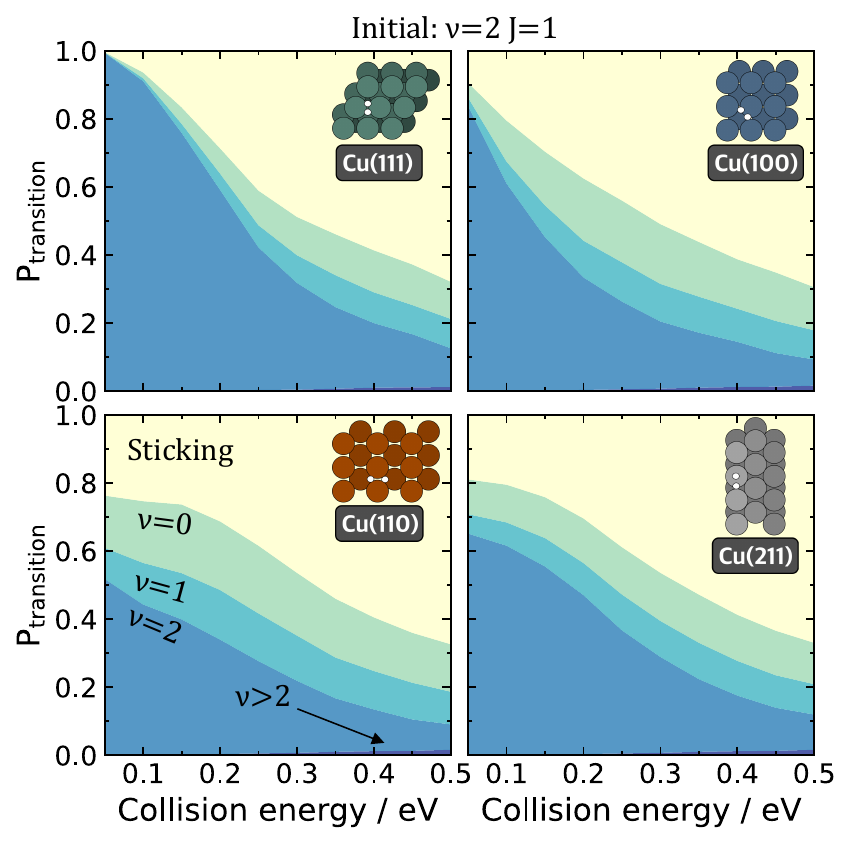}
            \caption{\textbf{
            Transition probabilities as stacked area plots for H\textsubscript{2} scattering at Cu surfaces (300~K).} Final states after scattering of H\textsubscript{2}($\mathrm{\nu}$=2,~J=1) are labeled. All final rotational quantum numbers are included. The vibrational excited population ($\mathrm{\nu}>2$) is very small. The probabilities were calculated using MD.
            }
            \label{fig:md_contour}
        \end{figure}

        \begin{figure}
            \centering
            \includegraphics[width=0.6\linewidth]{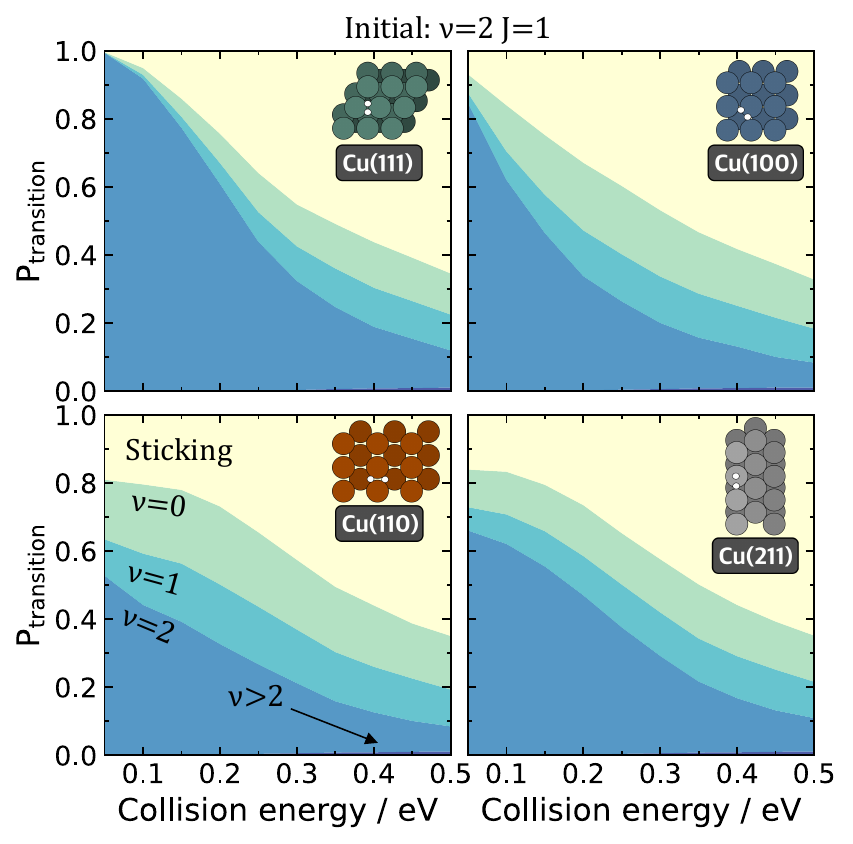}
            \caption{\textbf{
            Transition probabilities as stacked area plots for H\textsubscript{2} scattering at Cu surfaces (300~K).} Final states after scattering of H\textsubscript{2}($\mathrm{\nu}$=2,~J=1) are labeled. All final rotational quantum numbers are included. The vibrational excited population ($\mathrm{\nu}>2$) is very small. The probabilities were calculated using MDEF-LDFA.
            }
            \label{fig:ldfa_contour}
        \end{figure}

\clearpage